\documentclass[wrr]{agutex}
\usepackage[dvips]{graphicx}
\usepackage{gensymb}
\usepackage{amsmath}

\volume{51}
\yearofpublication{2015}
\articlenumber{017196}
\received{7 Mar 2015}
\revised{15 Oct 2015}
\accepted{21 Oct 2015}

\paperid{2015WR017196}

\authorrunninghead{MOURA, FIORENTINO ET AL.}

\titlerunninghead{BOUNDARY EFFECTS ON PRESSURE-SATURATION CURVES}

\authoraddr{Corresponding author: M. Moura,
Department of Physics, University of Oslo, Sem S\ae{}lands vei 24, 0371 Oslo, Norway.
\mbox{(marcel.moura@fys.uio.no)}}

\begin{document}

\title{Impact of sample geometry on the measurement of pressure-saturation curves: experiments and simulations}

\authors{M. Moura\altaffilmark{1}, E.-A. Fiorentino\altaffilmark{2}, K. J. M\aa{}l\o{}y\altaffilmark{1}, G. Sch\"afer\altaffilmark{3} and R. Toussaint\altaffilmark{2}}

\altaffiltext{1}{Department of Physics, University of Oslo, Norway.}
\altaffiltext{2}{Institut de Physique du Globe de Strasbourg, University of Strasbourg, France.}
\altaffiltext{3}{Laboratoire d'Hydrologie et de G\'eochimie de Strasbourg, University of Strasbourg, France.}

\begin{abstract}
In this paper we study the influence of sample geometry on the measurement of pressure-saturation relationships, by analyzing the drainage of a two-phase flow from a quasi-2D random porous medium. The medium is transparent, which allows for the direct visualization of the invasion pattern during flow, and is initially saturated with a viscous liquid (a dyed glycerol-water mix). As the pressure in the liquid is gradually reduced, air penetrates from an open inlet, displacing the liquid which leaves the system from an outlet on the opposite side. Pressure measurements and images of the flow are recorded and the pressure-saturation relationship is computed. We show that this relationship depends on the system size and aspect ratio. The effects of the system's boundaries on this relationship are measured experimentally and compared with simulations produced using an invasion percolation algorithm. The pressure build up at the beginning and end of the invasion process are particularly affected by the boundaries of the system whereas at the central part of the model (when the air front progresses far from these boundaries), the invasion happens at a statistically constant capillary pressure. These observations have led us to propose a much simplified pressure-saturation relationship, valid for systems that are large enough such that the invasion is not influenced by boundary effects. The properties of this relationship depend on the capillary pressure thresholds distribution, sample dimensions and average pore connectivity and its applications may be of particular interest for simulations of two-phase flow in large porous media.
\end{abstract}

\begin{article}

\section{Introduction\label{int}}
Fluid flow in porous media is an ubiquitous phenomenon. The subject is important to a broad class of professionals, ranging from engineers interested in increasing the recovery rates of oil reservoirs, to environmentalists concerned about the possible damages associated with liquid waste disposal and even to baristas wishing to make the perfect cup of espresso. Whether intended to improve the economy, protect nature or please the senses, the study of fluid dynamics inside a porous medium has received considerable attention from the scientific community. The practical relevance of the subject was long ago highlighted in (and motivating to) the fundamental experimental work of Henry Darcy \citep{darcy1856,brown2002}, where the connection between the flow rate inside a porous sample and the imposed pressure head difference was first described.  In addition to its inherent practical importance, interest in the theoretical aspects of porous media flow phenomena has also led the physics community to extensively analyze (theoretically, numerically and experimentally) the morphology of the phenomenon \citep{maloy1985,lenormand1985,lenormand1988,lenormand1989} and its dynamics \citep{furuberg1988,maloy1992,furuberg1996}. Those studies address the problem at various length scales, from the microscopic (pore-level) to the macroscopic (reservoir scale, for example) and different numerical/analytical tools and experimental techniques are employed in each of them \citep{grunde2005,toussaint2012,erpelding2013}.

Two-phase flow is encountered widely in the fields of hydrology, among other reasons, because soil and groundwater pollution by Dense Non-Aqueous Phase Liquids (DNAPL), such as chlorinated solvents (e.g., trichloroethylene (TCE)) constitute a large and serious environmental problem \citep{cohen1993}. Identification of pollution sources is difficult due to the fact that organic pollutants can rapidly migrate down to the bottom of the aquifer and/or along flow paths that differ from the water \citep{jellali2001,bohy2006,dridi2009,nsir2012}. In addition, in both soil and groundwater, they are subject to natural attenuation.

When a non-wetting low-viscosity fluid displaces a wetting high-viscosity one in a porous medium, the displacing fluid tends to channel through the paths of lower flow resistance, thereby forming pronounced fingers, which evolve in a branching structure inside the porous network leaving a characteristic macroscopic invasion pattern \citep{maloy1985,chen1985,lenormand1989}. The formation of such patterns in porous media flow is a long-ranged consequence of the random nature of the porous network and the forces governing the dynamics at the microscopic scale (pore level). Capillary, viscous and gravitational forces are typically the main contributors to such dynamics \citep{birovljev1991,grunde2005,toussaint2005,toussaint2012}, and their interplay is usually characterized by a set of dimensionless parameters such as the Bond number (ratio of gravitational to capillary forces) and the capillary number (ratio of viscous to capillary forces). Together with some fluid dependent properties, such as the wettability, viscosity and density ratios (in the case of two-phase flow), these parameters determine the typical flow regime (stable displacement, capillary fingering or viscous fingering) \citep{lenormand1989,lenormand1989.2} with immediate consequences for the macroscopic transport properties. 

While the efforts of the physics community have been mostly directed towards characterizing and understanding displacement patterns and local flow properties, hydrogeologists and soil scientists, on the other hand, have studied such systems with the goal of finding empirical laws relating saturation and capillary pressure at the Darcy scale, a meso-scale in which the medium and the flow are described by continuous mathematical fields. The description of multiphase flows at this mesoscopic scale requires the use of two constitutive relationships: relative permeabilities as function of saturation, $k_r=f(S)$ and capillary pressure as function of saturation, $p_c=f(S)$ (in this paper called a pressure-saturation curve, or \mbox{P-S} curve, but also known as water retention curve or moisture content curve). The saturation $S$ of a given phase is understood as the ratio between the volume occupied by that phase in the porous medium and the total pore volume. The capillary pressure $p_c$ is defined as the difference between the pressures of each phase, i.e., for two-phase flow,

\begin{equation}
p_c = p_{nw}-p_w \:,
\label{capillary}
\end{equation}
where $p_{nw}$ and $p_w$ are respectively the pressures of the non-wetting and wetting phases.

Direct measurements of the relation $k_r=f(S)$ for unsaturated porous media are challenging to perform and it is frequently preferred to estimate this quantity using numerical models, such as the van Genuchten-Mualem model \citep{mualem1976}. This model uses the relation $p_c=f(S)$ as an input and it is known to be particularly sensitive to the shape of this pressure-saturation curve \citep{vogel1988,vogel2000}. As a consequence, simulations using the van Genuchten-Mualem parameters may have both their stability and final results affected by differences in the relation $p_c=f(S)$, as shown by \citep{ippisch2006}. Additionally, in simulations of fluid flow in porous media, particularly in the ones that require the integration of the extension of Darcy equations to multiphase flows \citep{scheidegger1974}, the relation $p_c=f(S)$ is also necessary to give closure to the total set of equations governing the dynamics. Because of these and related aspects, the problem of understanding the functional behavior of the pressure-saturation curve has attracted a considerable attention over the years (for further aspects regarding finite capillary numbers and dynamic effects, see \citep{grunde2011}).

The extension of Darcy equations used in multiphase flows form a highly non-linear system, due to the nature of the permeability and capillary pressure functions needed to give closure to the equations. Numerical treatment of this system of equations is discussed extensively in the literature \citep{ewing1984,durlofsky1993,helmig1997,nayagum2004,dichiara2010}. Alternative forms of governing flow equations have been investigated, such as the fractional flow formulation, where the immiscible displacement of air and water can be expressed in terms of two coupled equations, a mean pressure (or global pressure) equation, and a saturation equation \citep{chavent1986,binning1999}.

Two models are typically used to describe the capillary pressure-saturation relationship: the Brooks-Corey model \citep{brookscorey1964} and the van Genuchten model \citep{vangenuchten1980}. The capillary pressure-saturation relationship of Brooks-Corey is given by:

\begin{equation}
p_c=p_dS_e^{-1/\lambda} \:,
\label{eq:BC}
\end{equation}
where $S_e\equiv\frac{S_w-S_{wr}}{1-S_{nwr}-S_{wr}}$ is the effective saturation of the wetting phase (water), $S_w$ and $S_{wr}$ are respectively the saturation and residual saturation of the wetting phase, $S_{nwr}$ is the residual saturation of the non-wetting phase (air), $\lambda[-]$ is the so-called pore size index and $p_d[Pa]$ is the entry pressure which corresponds to the minimum pressure needed by the air phase to displace the water phase. The capillary pressure-saturation relationship of van Genuchten is expressed by:

\begin{equation}
p_c=\frac{1}{\alpha}\left[S_e^\frac{n}{1-n}-1\right]^\frac{1}{n} \:,
\label{eq:VG}
\end{equation}
for $p_c>0$, where $\alpha[Pa^{-1}]$ and $n[-]$ are the so-called van Genuchten parameters.

In this paper, we will focus on the experimental and numerical quantification of the capillary pressure-saturation curve, $p_c=f(S_{nw})$, in a synthetic porous medium. We have decided to measure this quantity as a function of $S_{nw}$, the saturation of the non-wetting phase (air) for convenience, as opposed to water retention curves which are usually measured as function of the saturation of the wetting phase $S_w=1-S_{nw}$. We will report results of experiments and simulations on two-phase flow in a quasi-2D porous network with quenched disorder, i.e., having a static random distribution of pore throats. We will focus on the importance of boundary effects on the flow patterns and on how these effects can lead to changes in the measurement of $p_c=f(S_{nw})$. It will be shown that the curvature of the pressure saturation curve towards zero saturation and towards final saturation are essentially due to boundary effects associated with two main processes: around the inlet, this curvature reflects how an initially flat air-liquid interface gradually acquires the morphology of the boundary of a capillary invasion cluster. Around the outlet, the curvature is associated to the gradual pressure buildup in response to the invasion of narrower pores in the vicinity of an outlet filter (a semipermeable membrane).

We want to stress out here that the aforementioned boundary effects may bring unphysical features to the \mbox{P-S} curves (due to the so-called ``end effects'' or ``capillary end effects'', mentioned on \citep[p.\,453]{bear1972}) and, more importantly, that similar effects may also be present in the results of core sample tests (with 3D natural porous media) widely employed in the field, such as the ones produced using the ``porous diaphragm method'' \citep{bear1972,dullien1979}. In this technique, the core sample is placed in contact with a porous diaphragm, a semi-permeable membrane that plays a similar role to the filter at the outlet of our experiments, letting the liquid (wetting phase) to pass through but not the air (non-wetting phase). The use of such outlet membranes may artificially induce the invasion of narrower pores in its vicinity that would not be reached in an unbounded porous medium. The water-retention curves resulting from such tests (analogue to the \mbox{P-S} curves in the current work) are usually fitted to parametric equations such as the Brooks-Corey and van Genuchten models Eqs.\,\ref{eq:BC} and \ref{eq:VG} \citep{brookscorey1964,vangenuchten1980}. We make the claim that the result of these data fits may be considerably affected by the unphysical boundary effects discussed, very particularly in the initial and final phases of the porous medium invasion (where the invasion front is close to the inlet and outlet boundaries). Since the \mbox{P-S} curve is used as a closure relation for the integration of two-phase flow equations to model situations where each REV is directly connected to its neighbors (and not separated from them by semi permeable membranes), care must be taken into assuring that the features of such curves are realistic and not influenced by the experimental apparatus employed to obtain them.

Significant deviations in the measurements of capillary pressures in 3D systems with and without filters have been reported previously \citep{oung2003,bottero2011}. The exact shape of the \mbox{P-S} curve at the entire system scale is dependent on the relative sizes of the inlet and outlet boundaries with respect to the system size. However, in real underground systems in a reservoir simulation, no semipermeable membranes are present between adjacent representative elementary volumes (REV), and the interface between the two fluids does not particularly adopt a planar shape when it reaches the virtual boundaries between these REV. For the purpose of obtaining relationships valid in large scale reservoir simulations, and not relationships characteristic of finite size artifacts in the measurement tests, we will thus focus on extracting from the experimental and numerical results, a behavior representative of flow in natural conditions, without such semipermeable or straight faces boundary conditions. By analyzing the \mbox{P-S} curves away from the boundaries (a situation representative of flow in the bulk of a porous medium) we show that the \mbox{P-S} curve reduces to a simplified step-like function, characterized by a single entrance threshold (a statistically constant macroscopic capillary pressure)  and a final saturation. The two characteristic  values of this simple behavior are entirely determined analytically: the macroscopic capillary pressure at non zero saturation is computed directly from the capillary pressure thresholds distribution of the pore-throats and the average pore connectivity, and the final saturation is shown to depend only on the REV scale chosen, in direct consequence of the fractal nature of the capillary invasion air cluster.

\section{Methodology\label{sec2}}

\begin{figure}[t]
	\centering
	\includegraphics[width=\linewidth]{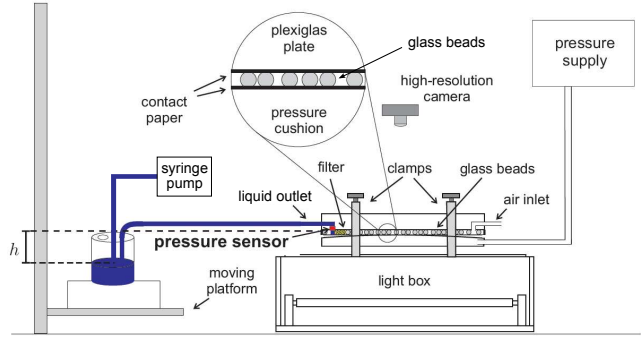}
	\caption{Diagram of the experimental setup. Additional details in appendix \ref{apxA}.}
	\label{fig:diagram}
\end{figure}

\subsection{Description of the experiment}
\subsubsection{Experimental setup\label{sec2.1}}
Real-time 3D experiments of flow in porous media are challenging to perform due to the natural difficulty involved in the visualization of the flow. Only very recently real-time pore-scale events in a porous rock were directly imaged, using high-speed microtomography \citep{berg2013}. The use of quasi-2D porous networks allows for a simplified experimental setup with the benefit of immediate visualization of the flow structures and real-time dynamics. A series of experimental studies has been performed in such systems using Hele-Shaw cells \citep{heleshaw1898} and variants to ensure the quasi-2D geometry of the flow \citep{maloy1985,meheust2002,grunde2005,toussaint2005}.

Fig.\,\ref{fig:diagram} shows a diagram of the experimental setup. The porous matrix employed is composed by a single layer of glass beads having diameters \mbox{$\O{}$} in the range \mbox{$1mm<\O{}<1.18mm$} that are placed onto the sticky side of a contact paper sheet. Silicon glue is used to define the in-plane boundaries of the model and another contact paper sheet is placed on top of it (with the sticky side facing the beads). The essential geometry of the model is thus defined by the silicon boundary (around it) and the contact paper sheets (above and below). In order to give robustness to the system and ensure that the quasi-2D geometry will be kept during the experiment, the porous matrix formed is pushed from below by a pressure cushion against a rigid Plexiglas plate placed on the top. The cushion ensures that the beads will be kept in place, despite the fluctuations in bead diameters. The Plexiglas plate has been previously milled to define channels for the inlets and outlets. Cuts were made in the upper contact paper sheet such that the liquid can be injected into and withdrawn from the porous network through the channels in the Plexiglas plate. In the experiments performed, the porous matrix was initially filled with a wetting viscous liquid composed of a mixture of glycerol ($80\%$ in weight) and water ($20\%$ in weight) whose kinematic viscosity and density were measured to be, respectively, $\nu = 4.25\;10^{-5} m^2/s$ and $\rho = 1.205\;g/cm^3$. The tabulated value for the surface tension at $25\degree$ C is  $\gamma = 0.064 \;N.m^{-1}$ \citep{gl1963}. Since the medium is initially completely wet by the liquid, the contact angle is always found to assume low values, although its exact value varies during a pore invasion due to dynamical effects.

The system also includes a filter placed in between the glass beads and the liquid outlet, at the end of the porous network. The filter is made of a sponge with pores much smaller than the typical pore size of the network itself and is intended to mimic the function of similar filters present in the apparatus typically used in the determination of \mbox{P-S} curves for core rock and soil samples (for example, in the so-called ``porous diaphragm method'' \citep{bear1972,dullien1979}).

The cell is positioned horizontally. Two pressure sensors (Honeywell 26PCAFG6G) were positioned at the outlet of the model and images were recorded from the top by a digital camera (NIKON D7100). A thermistor was placed in the liquid phase, in order to monitor its temperature. The air saturation is obtained from the analysis of the images. The outlet of the model is connected via tubing to a liquid reservoir which is placed on a translation stage (see Fig.\,\ref{fig:diagram}). The stage is controlled by a step motor which allows for precise changes in the height of the liquid reservoir. Since the liquid in the reservoir is connected to the liquid in the porous medium, by changing the height of the reservoir one can effectively control the pressure in the liquid phase inside the porous network. During the course of an experimental run the position of the reservoir is lowered, which reduces the pressure in the liquid initially saturating the porous medium. The model inlets are open to the atmosphere and, once the pressure in the liquid phase is low enough, the capillary pressure (see Eq.\,\ref{capillary}) becomes sufficiently large to overcome the minimum threshold value that is needed to invade the largest pore-throat available to the air-liquid interface. The air phase then starts to displace the liquid phase, initiating the invasion process. The capillary pressure threshold value is defined by the local pore geometry and the fluids wetting properties. It corresponds to the minimum difference in pressure between the phases necessary to drive the invasion of a given pore-throat. In accordance with Young-Laplace law \citep{bear1972}, larger pore-throats have lower values of capillary pressure thresholds and are, therefore, invaded earlier.

We are interested in studying flow in the capillary regime, in which the capillary forces dominate over viscous ones. Therefore, the fluid invasion must happen very slowly in order to avoid viscous effects. The typical duration of each experiment is more than $3$ days, and, for such long experiments, evaporation effects may play an important role and they had to be taken into account in the design of the experiment. Very particularly, the evaporation of liquid from the external reservoir had to be considered, since that could change the imposed pressure difference in the long-run (evaporation increases the height difference $h$ between the porous model and the liquid level in the reservoir, thus increasing the imposed pressure difference). Another effect that had to be taken into account was the fact that the displaced liquid volume that leaves the porous medium also goes into the liquid reservoir (decreasing the height difference $h$ and thus decreasing the imposed pressure difference). This second effect could possibly have been easily solved by using a reservoir large enough to make the change in the liquid level negligible during the course of an experiment. Nevertheless increasing the cross-sectional area of the reservoir does not solve the first issue because the evaporation rate also increases with this area. The solution that best suited us to handle both issues was to construct an overflowing mechanism into a specially designed liquid reservoir, which we describe in details in appendix \ref{apxA}. 

Fig.\,\ref{fig:evolution} shows a typical invasion process. The time difference between the first and last images is about 82 hours. The air inlet is on the left and the filter is the black stripe on the right. The time and instantaneous capillary pressure corresponding to each snapshot are shown under each image. Fig.\,\ref{fig:invasion} shows a spatiotemporal evolution of the whole invasion process. The colormap indicates the time in hours. 

\begin{figure}
	\centering
	\includegraphics[width=\linewidth]{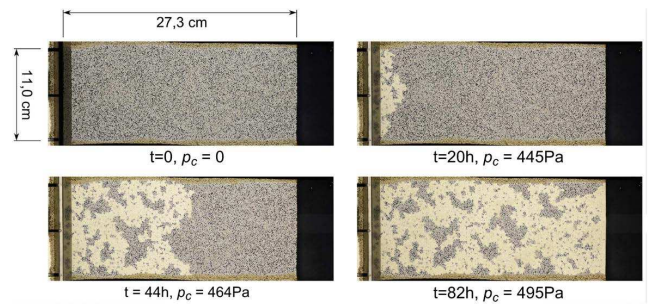}
	\caption{Images from the drainage process. The time $t$ of each image is shown in hours and the instantaneous capillary pressure $p_c$ in pascal. The porous medium is positioned horizontally to avoid gravitational effects. It is initially saturated with the liquid phase (blue) which is then gradually displaced by the invading air phase (white).}
	\label{fig:evolution}
\end{figure}

\begin{figure}
	\centering
	\includegraphics[width=\linewidth]{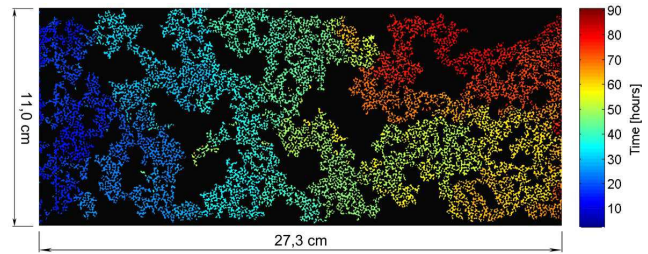}
	\caption{Experimentally measured spatiotemporal evolution of the drainage process for the same sample and experimental run as shown in Fig.\,\ref{fig:evolution}. The colormap corresponds to the invasion time in hours.}
	\label{fig:invasion}
\end{figure}

\subsubsection{Control of the experiments}
\begin{figure}[t]
	\centering
	\includegraphics[width=\linewidth]{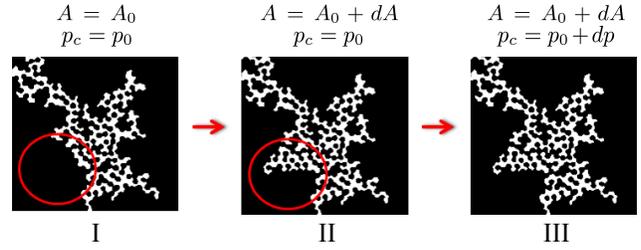}
	\caption{Feedback loop used to control the imposed capillary pressure in the experiments. The figure shows a detail from the black and white thresholded image, where the air phase is white and the liquid phase together with the porous medium are black. From I to II, the area of the air phase has grown from $A_0$ to $A_0+dA$, so the capillary pressure is kept constant at $p_0$ (no change in the liquid reservoir position). From II to III, the area of the air phase does not change, which means the capillary pressure must be increased from $p_0$ to $p_0+dp$ (by lowering the liquid reservoir and thus decreasing the pressure in the liquid phase by a positive amount $dp$). This analysis is done ``on the fly'' as the experiment is performed.}
	\label{fig:feedback}
\end{figure}

A computer running a Python code controls the experiment. A Keythley 2000 multimeter connected to the computer reads the signals from the pressure sensors and thermistor. One of the objectives in mind in designing the current experiment was to probe a response as quasistatic as possible, i.e., to be able to control the imposed pressure (by lowering the liquid reservoir) in such a manner that the capillary pressure (pressure difference between the air and liquid phases) would be just slightly above the lowest capillary pressure threshold defined by the geometry of the largest pore-throat available to the air-liquid interface. By doing so, one reduces the interference of dynamical viscous effects in the measurements (the \mbox{P-S} curves are known to be influenced by these effects \citep{grunde2011}). In order to achieve this, a feedback control loop was designed in which the imposed pressure in the liquid phase is decreased (thus increasing the capillary pressure) only when the system reaches an equilibrium configuration, i.e., when no pore is invaded for a certain amount of time in the whole model. Therefore, the code employed not only sends commands to the multimeter (to read pressure and temperature), step motor (to lower the liquid reservoir position) and camera (to capture images) but does so in an integrated manner, in which the current state of the experiment is used as an input to decide the next action. The procedure is as follows: pressure and temperature measurements are taken continuously and pictures are taken at fixed time intervals (approx. 15 seconds). At each 5th picture taken (5th, 10th, 15th...) the computer performs the image analysis to check whether the system is in equilibrium or not, i.e., to check if air is invading new pores or not. If the invasion process is still going on, the area of the air phase in, say, picture number 15 will be larger than the area in picture number 10, and the imposed pressure is left unchanged (no action is taken with respect to lowering the liquid reservoir). Conversely, if the area does not increase, the system is assumed to be in equilibrium and the imposed pressure difference is increased by a small amount (the liquid reservoir is lowered), see Fig.\,\ref{fig:feedback}. The amount $dh$ by which the reservoir is lowered is one of our control parameters. During the initial phase of the experiment, the liquid level in the reservoir is at the same height as the porous medium and the system needs to build up a considerable pressure until the invasion starts, therefore, we have chosen a larger $dh = dh_{max}$ for this initial process and a smaller $dh = dh_{min}$ to be used after the invasion has started (after the invading phase area reaches a certain small threshold). The values used were $dh_{max} = 0.01\;cm$ and $dh_{min} = 0.001\;cm$ corresponding to respective increments in the imposed capillary pressure of approximately $dp_{max} = 1.2\;Pa$ and $dp_{min} = 0.12\;Pa$.

\subsection{Image-based estimation of the capillary pressure thresholds distribution} \label{sec:voronoi}
\begin{figure}
	\centering
	\includegraphics[width=\linewidth]{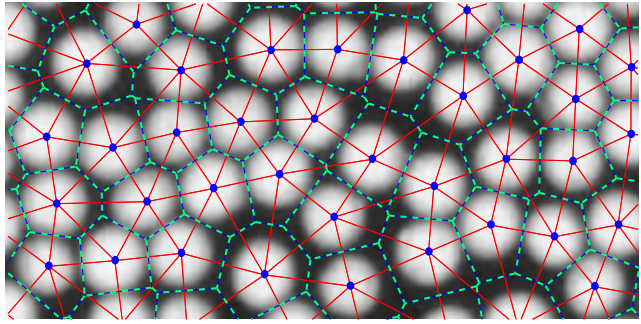}
	\caption{Pore-scale analysis used in the experimental determination of the capillary pressure thresholds distribution. Blue dots indicate the center of the beads, solid red lines show the Delaunay triangulation and the dashed green lines are the Voronoi lattice (dual graph to the Delaunay triangulation). The vertices of the Voronoi lattice mark the centers of the pores and the points at which the lines from the Delaunay triangulation and Voronoi lattice intersect are the centers of the pore-throats. The bead diameter is ca. $1 mm$.}
	\label{fig:voronoi}
\end{figure}

The use of a high-resolution camera allows us to analyze the images down to the pore scale. We have used the first image (in which the porous network is completely saturated with the liquid) to experimentally access the pore-throat size distribution. Initially we have identified the position and average radius of each bead in the network, next we employed a Delaunay triangulation algorithm \citep{lee1980} to find the distance between the centers of consecutive beads. By excluding the radius of each of the beads from the vector with all the pairwise distances between consecutive bead centers, we can directly access the size distribution of pore-throats. Fig.\,\ref{fig:voronoi} shows a zoomed in image of the procedure employed. The Voronoi lattice, (dual graph of the Delaunay triangulation \citep{voronoi1908}) is also shown. Each cell in the Voronoi lattice identifies the set of points in space that are the closest to the point in the center of the cell. Those central points are, by construction, the center of the glass beads. The vertices of the Voronoi lattice can be understood as the centers of the pores and the points at which the lines from the Delaunay triangulation and Voronoi lattice intersect are the centers of the pore-throats.

\begin{figure}
	\centering
	\includegraphics[width=\linewidth]{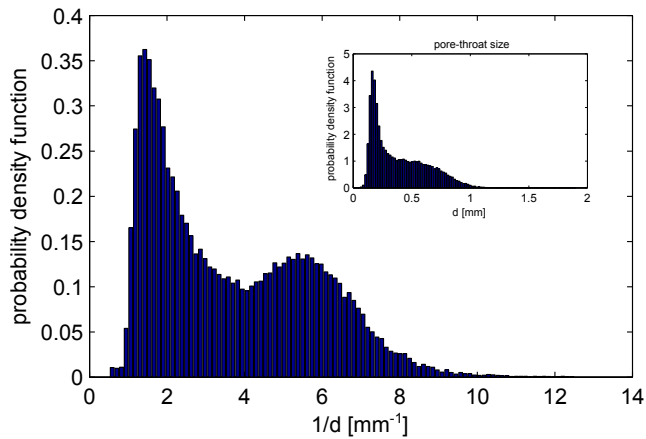}
	\caption{Image-based estimation of the capillary pressure thresholds distribution. The distribution of pore-throat sizes $d$ (in-plane throat diameters) is shown in the inset.}
	\label{fig:poresize}
\end{figure}

Fig.\,\ref{fig:poresize} shows the image-based estimation of the capillary pressure thresholds distribution. In the inset it is also shown the distribution of pore-throat sizes $d$, i.e., the in-plane throat diameters (the out-of-plane throat diameter is assumed to be constant and equal to $1mm$, the gap or height of the model). The capillary pressure thresholds distribution was estimated by considering the normalized histogram of the inverse of the pore-throat sizes $1/d$. In order to get to the actual capillary pressure distribution, it would be necessary to multiply $1/d$ by the surface tension $\gamma$ of the liquid-air interface (according to the Young-Laplace law) and correct it by i) a multiplicative geometrical factor, to account for the angle of contact between the air-liquid interface and the solid, dynamical wetting properties and local pore geometry and ii) an additive factor to account for the (constant) contribution of the out-of-plane curvature to the capillary pressure. As a first order approximation, we neglect such corrections here. They would lead to a rescaling and a translation of the horizontal axis in Fig.\,\ref{fig:poresize}, but would not significantly change the shape of the distribution.

\subsection{Description of the simulation procedure}

\subsubsection{Invasion percolation model}\label{sec:ip}
The simulation procedure is based on the invasion percolation model developed in \citep{wilkinson1983}. The idealized porous medium is a network of pores connected by throats. We work on a square lattice of sites oriented at 45\degree \! from the main flow direction: each site alternately representing the pore spaces or the rock, like in a chessboard. The length and width of the grid will be called respectively $l$ and $w$, and those will be expressed in number of lines and columns. To make the connection with the experiments clear, notice that the length is measured between the inlet and the filter at the outlet, whereas the width is measured in the direction perpendicular to this one.

Each pore is connected to its four closest neighboring pores by throats (also called necks, in other works, or bonds, in the percolation theory literature). The throat diameter fixes the capillary pressure threshold that is needed to make the injected fluid (air) jump from one site to the following through that connecting throat. Each time a site is invaded, its neighbors become available to the invading fluid. The capillary pressure threshold of a throat is randomly drawn from a uniform distribution inside the interval [$p_{min}$ $p_{max}$]. We chose $p_{min} = 200\;Pa$ and  $p_{max} = 608\;Pa$, giving $r_{max} = 0.64\;mm$ and $r_{min} = 0.21\;mm$ by the Young-Laplace law:
\begin{equation}
p_c = \frac{2 \gamma}{r_m}
\label{eq:YL}
\end{equation}
where $\gamma = 0.064 \;N.m^{-1}$ is the surface tension of the air-liquid interface \citep{gl1963} and $r_m$ the curvature radius of the duct.

Viscous forces are assumed to be negligible in comparison to capillary forces in the regime studied (low capillary numbers). At the initial state, the medium is completely filled with liquid, with the exception of the first row of pores that is filled with air. The displaced volume of liquid exits the network from one side of the lattice (opposite to the side where air enters the medium). A selection rule at this exit side mimics the effect of the filter in the experiments, blocking the passage of air but allowing the liquid to flow through.

Narrower pore-throats have larger capillary pressure thresholds and are, therefore, harder to invade. Since the invasion is governed by the values of capillary pressure thresholds associated to the throats (or bonds), this problem is partially similar to bond percolation \citep{stauffer1994}, although in that case, differently from the invasion percolation model, all throats with associated capillary pressure thresholds smaller than a given value would be invaded at once. In order to define which pore-throat will be invaded in the invasion percolation model, one has to perform a search in the available set of pore-throats (the ones that are at the liquid-air interface) and look for the throat with the lowest capillary pressure threshold $p_{ct}$. The pressure increment with respect to the instantaneous capillary pressure $p_c$ that is needed to invade this pore-throat is $\Delta p = p_{ct} - p_c$. Then, $p_c$ is incremented by $\Delta p$ and the pore connected through this throat is turned from the liquid state into the air state, making its neighboring pores available for the next step.

The choice of the invaded pore is also constrained by the incompressibility of the displaced liquid. When regions of the liquid happen to be completely surrounded by air or stuck between air and a wall, they become trapped, disconnected from the outlet, and these surrounded sites must be removed from the list of potential invasion candidates. Not incorporating this feature would lead, in the long term, to thoroughly filling the medium with air. The simulations stop when all the sites that are on the edge of the filter (exit side) contain air, blocking the flow of liquid outside of the model. The liquid left inside the porous network is found to be broken into a set of clusters that are disconnected from the exit side.

In the simulations, the two fluids are incompressible, so that the regular changes of pressure difference between the two phases equally represent a situation where air pressure is kept constant and water pressure is changed, or the contrary. In the experiments, in order to avoid effects due to air compressibility, the choice was made to reduce the pressure in the liquid instead of rising the air pressure.

\subsubsection{Capillary pressure thresholds: mapping between different distributions}\label{distribution}
We have chosen to work with a uniform distribution for the capillary pressure thresholds, but, due to the nature of the simulation procedure involved, we will show that drawing the pressure thresholds from another type of distribution would not change the geometry of invasion pattern -- although it would affect the \mbox{P-S} curve of the system. We will compare the outcome of the simulations that use a uniform distribution to the one produced by a distribution that was estimated empirically from the direct experimental measurement of the pore-throat sizes, as described in Sec.\,\ref{sec:voronoi}. In order to do so, we must first produce a mapping linking the capillary pressure thresholds withdrawn from the uniform distribution to the ones following the experimentally estimated distribution, as explained next.

In probability theory, the cumulative distribution function (CDF) $F(x)$ associated with a random variable $X$ is defined as the probability that $X<x$, i.e.,

\begin{equation}
F(x)=\int_{-\infty}^x f(x')dx' \:,
\label{cdf_def}
\end{equation}
where the random variable $X$ follows a distribution given by the probability density function (PDF) $f(x)$. In the context of percolation theory \citep{stauffer1994}, the CDF $F(x)$ is also known as the occupancy probability.

Let us call $F_{unif}(p_{unif})$ the CDF of the capillary pressure thresholds following a uniform distribution and $F_{emp}(p_{emp})$ the CDF of the capillary pressure thresholds following the empirically measured distribution. The set of capillary pressure thresholds uniformly distributed is generated using a standard random number generator, its PDF and CDF being given, respectively, by

\begin{equation}
f_{unif}(p_{unif})=\frac{1}{p_{max}-p_{min}} \ \ \textrm{and}
\label{funif_def}
\end{equation}

\begin{equation}
F_{unif}(p_{unif})=\frac{p_{unif}-p_{min}}{p_{max}-p_{min}} \:,
\label{Funif_def}
\end{equation}
where $p_{min}$ and $p_{max}$ are the minimum and maximum values attainable by the pressure. The capillary pressure thresholds distributed according to the empirically measured distribution are generated using a rejection sampling method, which is well described in the literature (see, for instance, \citep{press1992}). Since we do not have access to analytical forms for the empirical PDF $f_{emp}(p_{emp})$, we infer this quantity experimentally, as described in Sec.\,\ref{sec:voronoi} and shown in Fig.\,\ref{fig:poresize}, by normalization of the measured histogram of entry pressures (determined from the pore-throat geometries). The calculation of the empirical CDF $F_{emp}(p_{emp})$ is done, via numerical integration using Eq.\,\ref{cdf_def}.

As stated, the change in distribution of the capillary pressure thresholds does not affect the invasion pattern in the simulations. This happens because air follows the path of least resistance, which does not depend on the particular values of the pressure thresholds themselves, but only on their rank in the ordering from the smallest to the largest capillary pressure threshold. Such a rank remains unchanged by a change in the distribution. This property is used to define a correspondence between pressure values in the uniform and empirical distributions: since the rank is unchanged, $F_{unif}(p_{unif}) = F_{emp}(p_{emp})$, therefore $p_{emp} = F^{-1}_{emp}(F_{unif}(p_{unif}))$.

\section{Results and discussion}
\subsection{Measurements of \mbox{P-S} curves}

\subsubsection{Experiments}\label{sec:experiments}
\begin{figure}
	\centering
	\includegraphics[width=\linewidth]{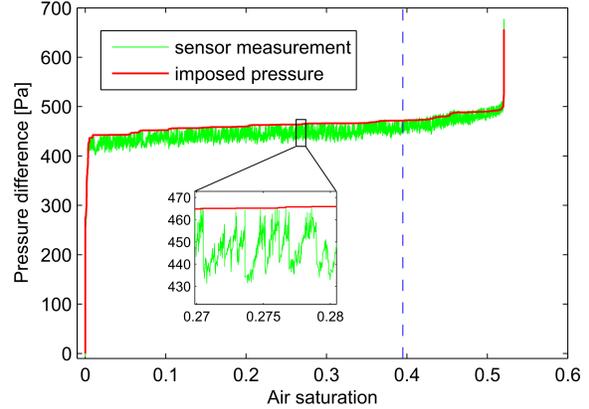}
	\caption{Experimental \mbox{P-S} curve obtained for the model depicted in Figs.\,\ref{fig:evolution} and \ref{fig:invasion}. The thick red curve is the imposed capillary pressure obtained by multiplying the height difference $h$ between the model and the liquid reservoir by the liquid's specific weight $\rho g = 11.831 kN/m^3$. The thin green curve shows direct pressure measurements at the outlet of the system made with a pressure sensor. The inset is a zoomed in section showing the viscous pressure drops that follow each pore invasion event. The dashed blue line indicates the breakthrough saturation $S_{B}=0.395$, i.e., the saturation at which the air phase first percolates through the model, reaching the filter at the outlet.}
	\label{fig:pscpexp3}
\end{figure}

Fig.\,\ref{fig:pscpexp3} shows an experimentally measured \mbox{P-S} curve for the model depicted in Figs.\,\ref{fig:evolution} and \ref{fig:invasion}. Two kinds of pressure measurements are seen in this figure. The thick red line represents the capillary pressure head imposed to the model obtained by multiplying the liquid's specific weight $\rho g = 11.831 kN/m^3$ by the height difference $h$ between the porous network and the liquid reservoir (see Fig.\,\ref{fig:diagram}). The thin green line corresponds to the pressure measured at the model's outlet using an electronic pressure sensor. In a static situation the two curves coincide. The difference seen between those curves is due to viscous pressure drops arising from dynamical effects during the pore invasion events: once the capillary pressure is large enough to overcome the capillary pressure threshold imposed by the largest pore-neck in contact with the front, the invasion of that pore takes place, with air displacing the filling fluid. The motion of the fluid sets in viscous stresses which lead to the pressure drops seen in the green curve. This dynamics happens in an avalanche manner during which the invasion of one pore may trigger the invasion of other pores in the neighborhood and it has been extensively studied in the literature \citep{haines1930,maloy1992,furuberg1996}. In the current paper we neglect these dynamical effects and focus only on the quasi-static situation. For the experiments reported next, we have decided to employ pressure measurements obtained from the height difference between the model and the liquid reservoir, like the one in the thick red curve in Fig.\,\ref{fig:pscpexp3}. One of the reasons that has led us to chose this approach is the fact that this method seems to be more closely related to the kind of procedure employed in the field tests, for example in the ``porous diaphragm method'' using core rock or soil samples \citep{bear1972,dullien1979}.

\begin{figure}
	\centering
	\includegraphics[width=\linewidth]{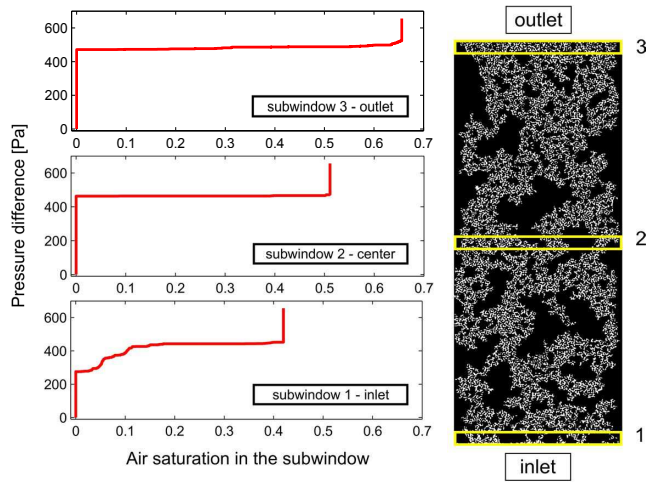}
	\caption{Experimental \mbox{P-S} curves obtained for 3 different measuring subwindows inside the model. The subwindows were chosen in order to emphasize the distinct effects caused by the boundaries on the curves. From bottom to top they correspond to the building regime (subwindow 1), propagation regime (subwindow 2) and clogging regime (subwindow 3).}
	\label{fig:subwindow}
\end{figure}

The properties of such pressure-saturation curves are certainly influenced by the characteristics of the porous medium, such as its porosity and local connectivity, but are in no sense uniquely determined by them. In fact, some features of the curves can be highly influenced by the effects of the boundaries of the model. The extent to which these boundary effects influence the overall \mbox{P-S} curve depends on the size and dimensionality of the porous medium itself. We have observed that the invasion close to the inlet and outlet of the model are particularly subjected to boundary effects. For the sake of clarity, we have decided to divide the flow into three parts: \textbf{building regime} (close to the inlet), \textbf{propagation regime} (central part of the model) and \textbf{clogging regime} (close to the outlet). In order to scrutinize in details the impact of the boundary effects in each regime, we have performed the analysis to produce \mbox{P-S} curves using measuring subwindows located inside the model in a set of regions either very close or very far from the boundaries. Fig.\,\ref{fig:subwindow} shows the resulting \mbox{P-S} curves for 3 of those subwindows each one corresponding to a particular invasion regime.

\begin{enumerate}
\item \textbf{Building regime:} in the subwindow close to the inlet, one can see the pressure building up phase as the invading front evolves from the initial flat interface to the ramified fractal profile characteristic of slow drainage processes in porous media \citep{lenormand1989.2,maloy1992}, see Fig.\,\ref{fig:evolution}. Initially the easiest (largest) pores are invaded and then, as the front progresses inside this subwindow, it is necessary to increase the capillary pressure to give access to harder (narrower) pores, since all the easy ones available to the front were already filled. This evolution progresses and is seen in the left part of the lower \mbox{P-S} curve (bottom curve in Fig.\,\ref{fig:subwindow}, subwindow 1 - inlet).
\item \textbf{Propagation regime:} after the invading front has progressed a certain characteristic distance inside the model, the imposed pressure becomes large enough, such that only minor increments from this value are necessary to keep driving the invasion through most of the central part of the model. By measuring the \mbox{P-S} curve using a subwindow in the middle of the porous network, one can thus see a much flatter profile (middle curve in Fig.\,\ref{fig:subwindow}, subwindow 2 - center). The regions of pressure building up are constrained to the extreme left and right of the curve, as vertical lines, which means they happen respectively before and after the invasion front has reached this subwindow. In the central portion of the model, where boundary effects are negligible, the air-liquid interface propagates steadily at a statistically constant capillary pressure $p_{crit}$. We will later characterize this critical value of the pressure and show how it depends on the medium and fluids' properties.
\item \textbf{Clogging regime:} for the subwindow close to the outlet, the presence of the filter forces the air phase to spread sideways, leading to an increase in the final air saturation inside this subwindow as can be seen in the upper \mbox{P-S} curve in the figure (top curve in Fig.\,\ref{fig:subwindow}, subwindow 3 - outlet). This happens because the typical pore sizes of the filter are much smaller than the pore sizes of the matrix, constraining the air invasion to remain inside the matrix. As the capillary pressure is increased, smaller pores become available and are invaded thus leading to the higher final saturation in this subwindow and the upwards curvature in the \mbox{P-S} curve. The invasion of these narrower pores in the vicinity of the filter is, therefore, a direct consequence of the presence of the filter itself and would not occur in an unbounded system. Once all the pores connected to the filter are invaded by air, the filter is said to be clogged and the liquid within it is disconnected from the porous medium. Increasing the pressure even further does not lead to any additional invasion as can be seen by the final vertical line in the \mbox{P-S} curve.
\end{enumerate}

In the following section we will make use of computer simulations to stress these boundary effects more clearly.

\subsubsection{Simulations}
\begin{figure}
\noindent\includegraphics[width=20pc]{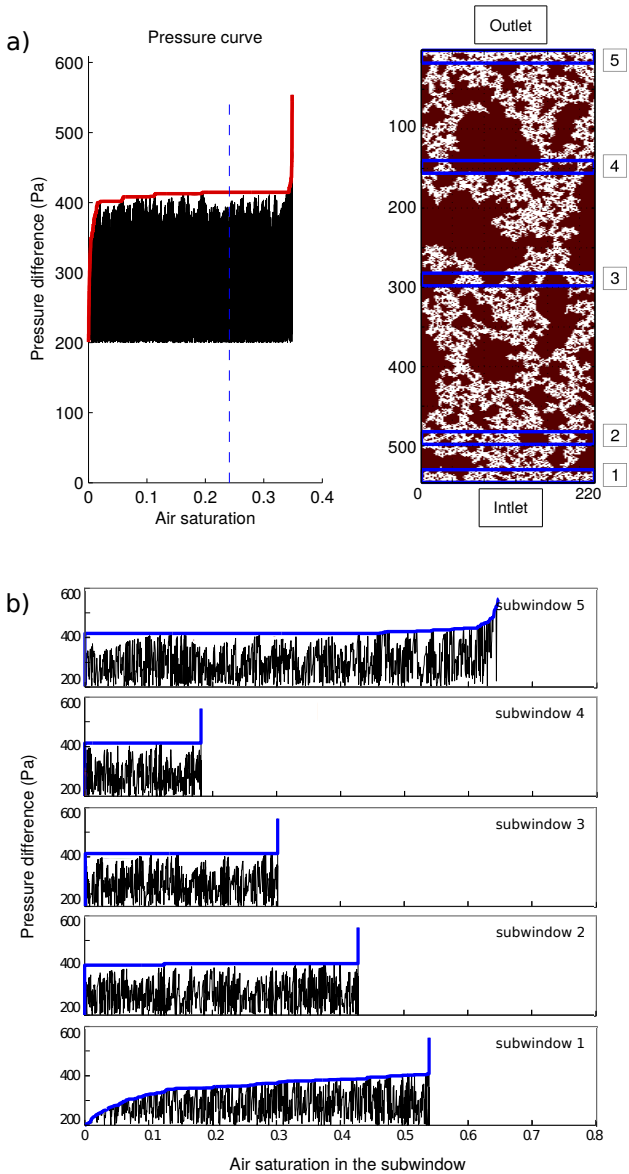}
\caption{a) \mbox{P-S} curve for the simulation of a system of size $w = 220$ columns and $l = 546$ lines. The capillary pressure thresholds values (drawn from a uniform distribution) are shown in black and the pressure envelope (\mbox{P-S} curve) in red. The breakthrough saturation $S_B$ is marked by the dashed blue line. The network image to the right shows air in white, liquid in brown, and the subwindows positions in blue. b) \mbox{P-S} curves measured inside the subwindows.}
\label{fig:subwindows}
\end{figure}

We have used the Invasion Percolation algorithm described in Sec.\,\ref{sec:ip} to simulate the invasion process. Assuming that a numerical site represents a $0.5mm$ size pore, we perform simulations with sizes similar to those of the experimental cell in Fig.\,\ref{fig:evolution} by defining a grid with dimensions $l=546$ lines and $w=220$ columns, representing, respectively, its length and width. The resulting \mbox{P-S} curve is shown on the left side of Fig.\,\ref{fig:subwindows}a). The black curve shows the successive values of capillary pressure thresholds of the invaded pores, drawn from a uniform distribution. Notice that this curve goes up and down intermittently, reflecting the fact that there is no imposed spatial correlation between the threshold values: after invading a given pore, the following pore may have a larger or lower capillary threshold. In the experimental situation, since the capillary pressure is set externally by the level of the liquid reservoir (see Fig.\,\ref{fig:diagram}), the invasion of an easier pore (larger, with lower pressure threshold) that immediately follows a harder one (narrower, with higher pressure threshold) happens in a sudden burst fashion, leaving the signature pressure drops depicted in the green curve in Fig.\,\ref{fig:pscpexp3}. As explained in Sec.\,\ref{sec:experiments}, in the experiments, we use the measurements from the imposed pressure, a strictly non-decreasing function. In order to simulate this quantity numerically, we define a pressure envelope curve, shown in red in Fig.\,\ref{fig:subwindows}a), which marks the value of the highest capillary pressure threshold reached so far, thus mimicking the behavior of the imposed pressure in the experiments. The continuous increase of this pressure envelope indicates the necessity of invading harder pores as air propagates through the medium. We have noticed in our simulations that the breakthrough marker (dashed blue line in Fig.\,\ref{fig:subwindows}a)) coincides with the inflexion point of the pressure envelope. This indicates that the presence of the external filter is responsible for the curvature inversion, i.e. that in an open medium, in the absence of any outside filter, there should be no upward increase of slope of this envelope. We remind that the breakthrough is defined as the moment in which the air phase first reaches the filter, forming a sample-spanning cluster of invaded pores \citep{stauffer1994}.

\begin{figure}
	\centering
	\includegraphics[width=\linewidth]{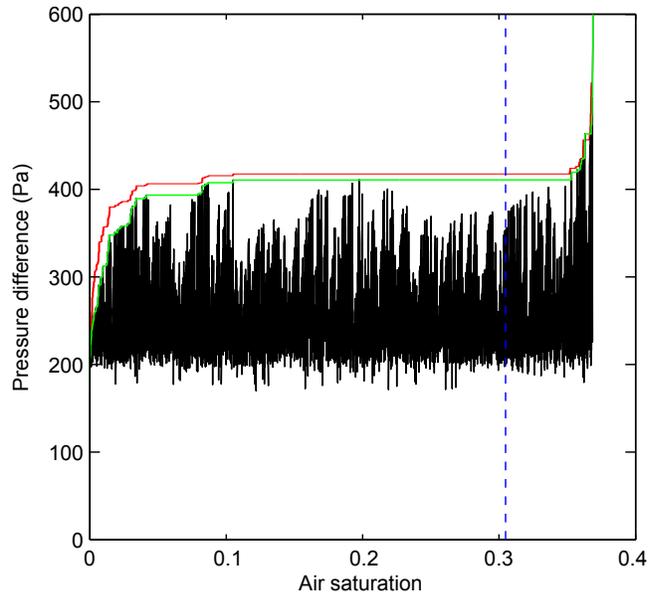}
	\caption{\mbox{P-S} curve (green line) simulated for a system of size $w = 220$ columns and $l = 273$ lines, having capillary pressure thresholds (black line) following the empirically measured distribution. The \mbox{P-S} curve obtained for a uniform distribution (red line) is shown for comparison. The amount by which these curves differ depends on how unequal the underlying distributions are. The breakthrough marker (blue dashed line) is the same for both curves, since the invasion order is not affected by the mapping between the distributions.}
	\label{fig:distrib_curves}
\end{figure}

Next, we have made use of the mapping technique discussed in Sec.\,\ref{distribution} to visually analyze the effects of the underlying capillary pressure thresholds distribution in the simulation of the \mbox{P-S} curves. Fig.\,\ref{fig:distrib_curves} shows the \mbox{P-S} curve (green line) associated with the empirically measured distribution from Sec.\,\ref{sec:voronoi}, together with the underlying pressure thresholds (black line). We compare this curve with another one where a uniform capillary pressure threshold distribution was used (red curve). We notice that, although the invasion pattern is left unchanged by the change in the distribution (due to the nature of the invasion percolation algorithm, as described in Sec.\,\ref{distribution}), the \mbox{P-S} curves are sensitive to this change (as they should in a realistic scenario). The amount by which they differ depends on how unequal the distributions used are. In producing the \mbox{P-S} curve for the empirically measured distribution, we have tuned the minimum and maximum values of the pressure, $p_{min}$ and $p_{max}$, to get results in the same order of magnitude as in the uniform case.

In order to emphasize the boundary effects to the \mbox{P-S} curves, we select a set of subwindows, as shown in the right part of Fig.\,\ref{fig:subwindows}a), and measure the \mbox{P-S} curves inside of them, leading to Fig.\,\ref{fig:subwindows}b) (in a similar fashion to what was done experimentally in Fig.\,\ref{fig:subwindow}). From the analysis of Fig.\,\ref{fig:subwindows}b), we see that the results from our simulations reinforce the experimental observations of the previous section. We can directly observe that the phases of initial pressure build up are restricted to the region close to the inlet, as shown by the rise in the pressure envelope in subwindow 1, where the air front evolves from the initial straight line to the capillary fingering ramified pattern, characterizing the \textbf{building regime}. The \textbf{clogging regime}, with the final rise in pressure due to the invasion of narrower pores in the vicinity of the filter at the outlet is shown in subwindow 5. For the subwindows placed far from the boundaries, we see the much flatter pressure envelope from the \textbf{propagation regime}, also in accordance with what was observed in the experiments, see for example, subwindow 3. The phases of pressure building up appear as vertical lines in the extreme left and right of the curve, meaning, as stated earlier, that they occur before and after the invading front has reached the subwindow area. In the interior of the subwindow, the invasion happens at an essentially constant capillary pressure, thus producing a flat plateau in the \mbox{P-S} curves for these areas. Our numerical results, supported by experimental confirmation, yield the conclusion that when boundary effects are negligible, the invasion happens at a statistically constant capillary pressure $p_{crit}$. In this region (that for a large enough sample would correspond to the majority of the flow), the \mbox{P-S} curve is reduced to a very simple function, needing only two values for its characterization: the critical value of the pressure $p_{crit}$ and the final saturation $S_F$. We proceed to a deeper numerical analysis of these quantities, starting from the latter.

\subsection{Analysis of the final air saturation}\label{sec:sat}
Let $S_F$ denote the final air saturation, corresponding to the air saturation at the end of the experiment, i.e., when all the sites connected to the exit side are invaded by air and no additional displacement is possible. Due to the fractal nature of the capillary invasion process \citep{mandelbrot1982,feder1988,lenormand1989.2}, the final saturation is expected to be a function of the system's geometry. In order to analyze this dependency, an extensive numerical study was performed in which the dimensions (length $l$ and width $w$, expressed respectively in number of lines and columns) were separately changed.

\begin{figure}
\noindent\includegraphics[width=20pc]{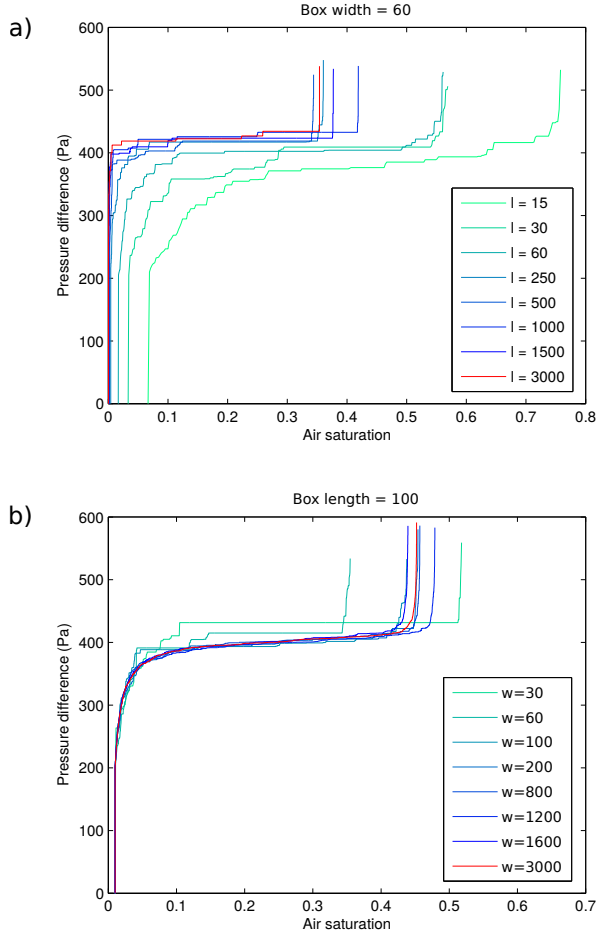}
\caption{Simulations of \mbox{P-S} curves for various system sizes. a) Curves for several lengths $l$ with width fixed to $w=60$ columns. b) Curves for several widths $w$ with length fixed to $l=100$ lines.} the
\label{fig:shape}
\end{figure}

Let us initially analyze the evolution of the \mbox{P-S} curves for different values of $l$ and $w$. In Fig.\,\ref{fig:shape}a), \mbox{P-S} curves are produced for systems with different lengths $l$ and same width $w=60$ (for the sake of clarity, we omit here the actual pressure threshold values and show only the pressure envelopes). We observe that for systems in which the length $l$ (distance between inlet and outlet) is too small, boundary effects dominate the whole invasion process: there is a tendency towards larger final saturations and the \mbox{P-S} curves do not reach a flat plateau. The initial curvature (related to the flow close to the inlet), seems to extend longer in the \mbox{P-S} curves than the final curvature (related to the flow close to the outlet). This is particularly noticeable in the systems with shorter lengths in Fig.\,\ref{fig:shape}a). Therefore, when it comes to the influence on the overall shape of the \mbox{P-S} curve, the boundary effects coming from the flow close to the inlet (building regime) are more important than those coming from the flow close to the outlet (clogging regime), and require a longer system length $l$ to fade out. As the system's length $l$ is increased, the final saturation is reduced, and the phases of pressure building up are restricted to the regions close to the inlet and outlet. For long enough systems, most of the invasion happens at an essentially constant capillary pressure (the propagation regime dominates), which is reflected as the plateau observed in the \mbox{P-S} curves for large $l$. Notice that for those long systems, the fraction of pores in the vicinity of the boundaries is small therefore their \mbox{P-S} curves resemble the ones in subwindow 3 in Fig.\,\ref{fig:subwindows} where the measurements are taken at the center of the model (far from the boundaries).

In Fig.\,\ref{fig:shape}b), \mbox{P-S} curves are produced for systems with different widths $w$ and same length $l=100$. One can observe in this figure the convergence of the final air saturation as the width is increased but, for small widths, the dispersion in the saturation does not seem to be as pronounced as in the case of small lengths (compare with Fig.\,\ref{fig:shape}a)). Putting it in a different way: samples with small distance $l$ between inlet and outlet seem to have their final air saturation more affected by boundary effects than samples with small distance $w$ between the lateral boundaries. One possible explanation for such result lies in the fact that the final saturation is increased in the vicinity of the outlet boundary, due to the invasion of narrower pores induced by the filter in that region. When the sample length is reduced, the relative importance of this region close to the outlet is increased, thus increasing the overall saturation as described previously. Taking these observations into account, we shall next analyze the behavior of the final saturation as the length $l$ of the system is gradually changed for a given set of fixed system's widths $w$. Additional analysis for the air saturation are included in appendix \ref{apxB}: the variation of the final saturation as the width $w$ of the system is gradually changed, for a given set of fixed system's lengths $l$, can be found in Sec.\:B1 and similar considerations regarding the dependence of the breakthrough saturation on the system geometry can be found in Secs.\:B2 and B3.

\subsubsection{Variation of the final air saturation with the system's length}\label{sec:sat_l}

\begin{figure}
\noindent\includegraphics[width=20pc]{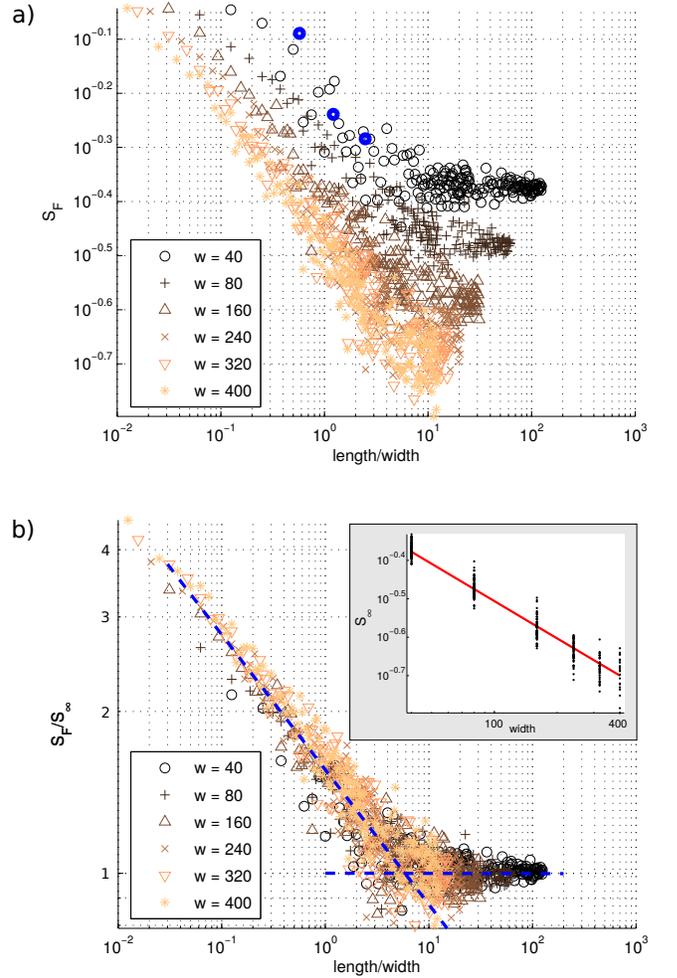}
\caption{Simulations. a) Final air saturations as a function of $l/w$ for several widths with $l$ varying from 5 to 5000. The blue points are from the experimental data in Fig.\,\ref{fig:pscomp3}. b) Same curves divided by $S_{\infty}$, the average value of $S_F$ at large $\frac{l}{w}$. The dashed blue line corresponds to the linear regression giving the exponent $D_c\simeq1.75\pm0.10$. The inset corresponds to $S_{\infty} \propto w^{(D_c^\infty-D)}$ with $D_c^\infty\simeq1.68\pm0.10$ and $D=2$.}
\label{fig:length_satFin}
\end{figure}

Fig.\,\ref{fig:length_satFin}a) shows an extensive numerical study produced to analyze the dependency of the final saturation on the system's length $l$. Each point in this figure is obtained by 1) fixing a geometry for the grid (length and width); 2) letting the invasion take place according to the invasion percolation algorithm described in Sec.\,\ref{sec:ip}; and 3) computing the final air saturation. For several widths $w$ (shown in the legend in the figure), the length $l$ of the grid is chosen in the interval from $l=5$ to $l=5000$, with a step of 5 from 5 to 100, 10 from 100 to 1000 and 50 from 1000 to 5000. For each fixed width, the final saturation as a function of the ratio $r=l/w$ is plotted in Fig.\,\ref{fig:length_satFin}a), different colors and symbols being used for different widths. We observe evidence of scaling behavior up to a certain crossover region after which the final saturation seems to remain constant. This observation leads us to postulate the scaling law 
\begin{equation}
S_F(r,w)=r^{(D_c-D)}f(r)S_{\infty}(w) \:,
\label{eq:Sf}
\end{equation}
where $D_c$ is the fractal dimension \citep{mandelbrot1982} of the invasion pattern, $D=2$ is the Euclidean geometrical dimension, $f(r)$ is a crossover function, and $S_{\infty}(w)$ is the final saturation for a model with width $w$ and length $l\rightarrow\infty$. The crossover function is defined as
\begin{equation}
f(r) =
\left\{
	\begin{array}{ll}
		C  & \mbox{for small } r \\
		r^{-(D_c-D)} & \mbox{for large } r 
	\end{array}
\right.
\label{eq:f}
\end{equation}
where $C$ is a dimensionless constant, numerically determined as $C\simeq 1.56$ by a linear regression in the bilogarithmic plot shown on Fig.\,\ref{fig:length_satFin}a), over the ensemble of simulations at $r<5$. In order to test our assumption for the functional dependency of $S_F$, we divide the data in Fig.\,\ref{fig:length_satFin}a) by the corresponding value of $S_{\infty}(w)$, determined by averaging simulation values of $S_F(r,w)$ at large r. The result is shown in Fig.\,\ref{fig:length_satFin}b). The data collapse shown in this figure indicates that $S_F/S_{\infty}$ is indeed a function of $r$ only and, therefore, Eqs.\,\ref{eq:Sf} and \ref{eq:f} seem to provide a reasonable description. The exponent $D_c$ in Eq.\,\ref{eq:Sf} was measured from Fig.\,\ref{fig:length_satFin}b) as the slope of the dashed linear regression at $r<5$, and found to be $D_c\simeq1.75\pm0.10$. The dependency of $S_{\infty}$ with $w$ was also studied and we estimate that $S_{\infty} \propto w^{(D_c^\infty-D)}$ with $D_c^\infty\simeq1.68\pm0.10$ and $D=2$, see the inset in Fig.\,\ref{fig:length_satFin}b). From Fig.\,\ref{fig:length_satFin}b) we can also give a more reasonable definition for the terms ``small $r$'' and ``large $r$'' used in Eq.\,\ref{eq:f}: from the collapsed data, we can say that $r\simeq5$ marks the approximate turning point for the crossover function (corresponding to the crossing point of the two asymptotic regimes shown in blue dashed lines). Therefore, we can say that in the case of our 2D model, the system's final saturation becomes roughly independent of its length when this length is more than 5 times larger than the system's width.

There is an intuitive reasoning behind such kind of behavior for the final saturation: far enough from the boundaries the system becomes, in a statistical sense, translationally invariant along the inlet-outlet direction. That is to say, there is no reason to believe that one portion of the system would present a saturation pattern different from another portion located a bit further (as long as these areas are far from the inlet and outlet boundaries). As the length $l$ of the system is increased, this translationally invariant area also increases and becomes dominant for $l$ larger than a given value ($5$ times the width in our case). Although the experiments and simulations presented here are performed in a 2D model, the intuitive reasoning just presented is also valid in 3D. The numerical value for the crossover might be different, but the overall behavior for the saturation must be similar, and its scaling with the sample's length must follow a rule like Eq.\,\ref{eq:Sf}.

\subsubsection{Direct comparison between simulations and experiments}
\begin{figure}
	\centering
	\includegraphics[width=\linewidth]{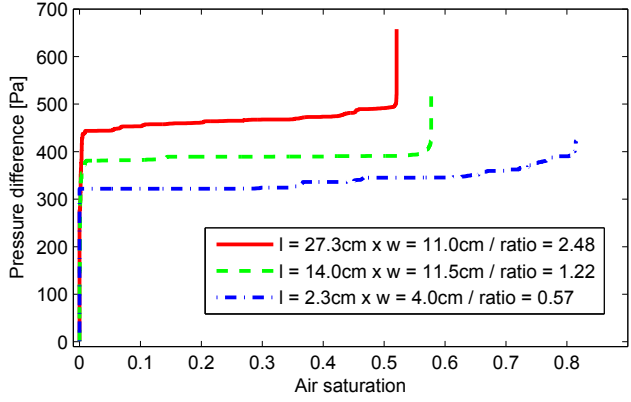}
	\caption{Experimental \mbox{P-S} curves for 3 different models (the models' geometries are stated in the legend). The lower curve extends longer to the right, indicating that for this model the residual saturation in air is much higher. This is due to the fact that the boundary effects are enhanced by the smaller height of the model, see text. Additionally, we notice that the values of capillary pressures driving the invasion are different for each model. This difference arises from the fact that the distribution of pore-throat sizes is not exactly the same for every model, since the porous medium is rebuilt for each case.}
	\label{fig:pscomp3}
\end{figure}

Fig.\,\ref{fig:pscomp3} shows the experimentally measured \mbox{P-S} curves for three different models (with dimensions shown in the legend of the figure). The number of model geometries tested experimentally is much lesser than the amount of numerical tests, due to the natural difficulties in having to physically reconstruct the models for each different geometry. Therefore, we cannot extract the functional dependency of the saturation with the model's dimensions from the experiments (our data set is not large enough). Nevertheless, from the analysis of this figure, one can at least say that experiments and simulations show similar trends: the residual saturation is influenced by the system size and, in particular, the model with smaller length presents a residual saturation in air much higher than the others, i.e., proportionally less liquid is left trapped inside the porous medium at the end of the experiment. The experimental values corresponding to the measurements in Fig.\,\ref{fig:pscomp3} are shown in Fig.\,\ref{fig:length_satFin}.a) as thick blue points.

We notice also that there are differences between the values of capillary pressures driving the invasion of the different systems shown in Fig.\,\ref{fig:pscomp3}. These differences arise from the fact that the distribution of pore-throat sizes may vary from one sample to another, since the porous medium is rebuilt for each experiment. The continuous red curve in Fig.\,\ref{fig:pscomp3} is the measured \mbox{P-S} curve corresponding to the longest model used, whose flow images are shown in Fig.\,\ref{fig:evolution} and having the pore-throat sizes distribution shown in the inset of Fig.\,\ref{fig:poresize}. Differences in pore-throat sizes are reflected in the respective capillary pressure thresholds and further translated into the critical value $p_{crit}$ of the capillary pressure necessary to drive the invasion, as will be discussed in Sec.\,\ref{sec:pcrit}. Additional details on the different capillary pressure thresholds distributions for the models in Fig.\,\ref{fig:pscomp3} are given in appendix \ref{apxC}.

\subsubsection{Final air saturation in square subwindows}

\begin{figure}
	\centering
	\includegraphics[width=\linewidth]{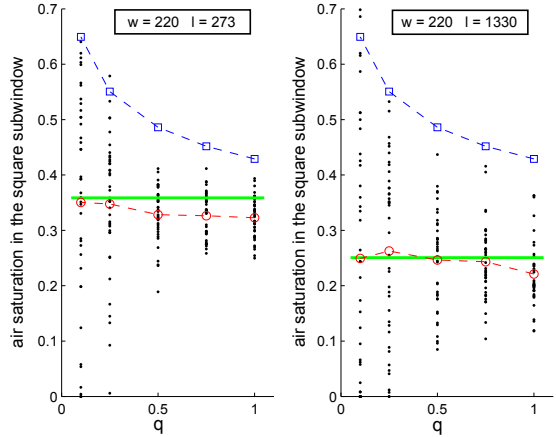}
	\caption{Final air saturation in the square subwindows of size $s$ as function of its relative size $q=s/w$ for 40 simulation processes in two different systems with width $w$ and length $l$ given in the legend. The blue line is the expected maximum air saturation in the subwindow, the red line is the mean final air saturation over all the processes in each subwindow and the green horizontal line is the average of the final air saturation in the whole system.}
	\label{fig:squares_middle_sat}
\end{figure}

Due to the fractal nature of the capillary invasion process, the air saturation depends on the size of the sample, as observed in Sec.\,\ref{sec:sat}. We can also notice a similar size dependency by measuring the saturation using subwindows of different sizes inside a given sample. In this section, we perform a numerical study considering square subwindows centered in the middle of the network, with sizes $s$ being a fraction $q$ of the system's width $w$: $s=qw$, where $q$ will take the values: $0.1$, $0.25$, $0.5$, $0.75$ and $1$ (the latter corresponding to having a square subwindow spanning the whole width of the grid). Fig.\,\ref{fig:squares_middle_sat} shows the outcome of these simulations for two different grid dimensions, the one on the left having width and length, respectively, $w=220$ and $l=273$ and the one on the right having $w=220$ and $l=1330$. The final saturations are represented by the black points, which are scattered around an average value $S_{ave}$ marked by a red circle for each subwindow size. This scattering is expected: it originates from the fact that the capillary pressure thresholds are drawn randomly at the beginning of each numerical experiment and, therefore, the invasion pattern differs from one realization to another. It is possible, for example, to have a measurement in which the saturation inside the subwindow is $S_F=0$, meaning simply that the invasion front has bypassed the area of that subwindow. The probability of this happening is of course higher for smaller subwindows. The blue squares represent the typical maximum saturation $S_{max}$ that can be expected for a system of size $s$, i.e., $S_{max}=\alpha s^{(D_c - D)}$. If the actual dimension of the square subwindow is $L$ (in, say, centimeters) and the typical distance between adjacent pores is $d$ (measured in the same length unit as $L$), $S_{max}$ is given by
\begin{equation}
S_{max} = \alpha \left(\frac{L}{d}\right)^{(D_c - D)} \:,
\label{eq:Smax}
\end{equation}
with $D_c=1.82$, the value of the fractal dimension of a capillary fingering pattern as in \citep{wilkinson1983}, and $\alpha=1.0$ for the data in Fig.\,\ref{fig:squares_middle_sat}. The green horizontal line corresponds to the average of the macroscopic final saturation $S_F$ in the whole system. We expect the average final saturation $S_{ave}$ inside a subwindow of a given size $s$ (red circles) to coincide with the average of the macroscopic final saturation $S_F$ (green line) in the limit of a very long system. This is expected, since the average saturation in the subwindows is an intensive quantity, i.e., not dependent on the system size. Considering that the macroscopic saturation is governed by the smallest dimension between the width and the length, we have

\begin{equation}
S_{ave} = \alpha \: \left(\frac{min(w,l)}{d}\right)^{(D_c - D)} \:.
\label{eq:Save}
\end{equation}
In the case of our simulations, we see that the final saturation $S_F$ (green line) is generally higher than the average final saturation $S_{ave}$ inside the subwindows (red circles) because of the more compact air invasion close to the boundaries of the model (which lie outside the subwindows).

\subsection{Analysis of the capillary pressure}\label{sec:pcrit}
As stated earlier, for a system that is long enough such that the influences of the boundaries to the flow are negligible, two parameters are essential to define the pressure-saturation relationship: the final saturation $S_F$ and the critical capillary pressure $p_{crit}$. We have analyzed the behavior of $S_F$ in the previous section and now we turn our attention to $p_{crit}$.

We start by defining a set of useful quantities. Let $y$ denote the distance from the inlet of the most advanced tip of the air front inside the model. If we think of the pressure as a function of this position, i.e.,
\begin{equation}
p_c=p_c(y) \:,
\label{eq:pcy}
\end{equation}
then, if the system's length is $l$, $p_c(l)$ is the pressure at breakthrough and $p_c(l/2)$ is the pressure when the invasion front is halfway through the model's length. Next, let
\begin{equation}
F^*=F(p_c)
\label{eq:Fstar}
\end{equation}
be the value of the cumulative distribution function $F$ associated with a given distribution of capillary pressure thresholds, calculated at the pressure $p_c$. We remind that $F^*$ varies in the interval $0 \leq F^* \leq 1$ and its numerical value corresponds to the probability of having a capillary pressure smaller than or equal to $p_c$ in the given distribution, see Sec.\,\ref{distribution}.
 
Let us focus now on the inverse of Eq.\,\ref{eq:pcy}, $y=y(p_c)$. This function gives the distance to the inlet of the most advanced tip of the air phase, as a function of the imposed capillary pressure. The plateau in the \mbox{P-S} curve is reflected in the curve $y=y(p_c)$ as a vertical asymptote at a given pressure, the critical pressure $p_{crit}$. When this pressure is reached the invasion of the air phase happens continuously making the tip position $y$ to advance inside the model. One could plot such function and evaluate $p_{crit}$ but such numerical value would be particular to the distribution of capillary pressure thresholds in use, and, therefore, uninteresting from a more general perspective. In order to overcome this, we have decided to consider instead the functional composition
\begin{equation}
y=y(p_c(F^*)) \:,
\label{eq:y}
\end{equation}
where we make use of the inverse of Eq.\,\ref{eq:Fstar}, to obtain the capillary pressure as a function of the CDF value, i.e., $p_c=F^{-1}(F^*)$. Due to the nature of the invasion percolation model, in which the invasion depends on the ranking of the capillary pressure threshold values and not on the values per se (see Sec.\,\ref{distribution}), the dependency of $y$ on $F^*$ is robust, in the sense that, apart from statistical fluctuations (that can be removed by averaging over several simulations), the curve $y(F^*)$ is independent of the particular capillary pressure thresholds distribution used and depends only on the connectivity of the underlying grid.

\begin{figure}
	\centering
	\includegraphics[width=\linewidth]{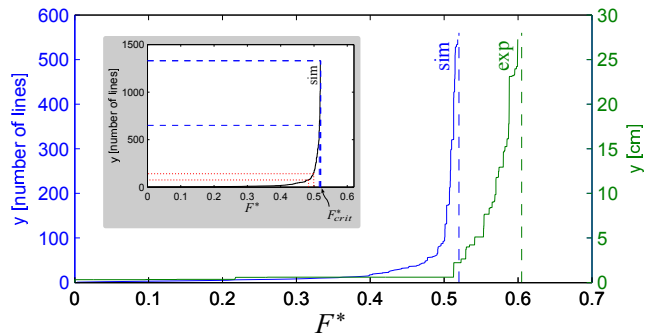}
	\caption{Maximum position $y$ of the air-liquid interface as function of the imposed capillary pressure $p_c$ expressed in terms of the CDF value via $F^*=F(p_c)$ (for universality regarding the type of pore-throat distribution). Results for simulations (blue, left) and experiments (green, right). The experimental measurements were made in the system shown in Fig.\,\ref{fig:evolution}. In the simulations, for each distance $y$ the corresponding applied pressure is obtained by averaging the result over 10 realizations.  The dimensions of the simulations were $w=220$ columns and $l=546$ lines in order to keep the same proportions of the experimental cell, having $w=11.0cm$ and $l=27.3cm$. We observe in both cases the diverging behavior of maximum air front position $y$ as the pressure CDF approaches the critical value $F^*_{crit}$, measured to be $F^*_{crit}=0.52$ in the simulations (connectivity 4) and $F^*_{crit} = 0.605$ in the experiments (average connectivity between 3 and 4). In the inset we show simulations made in order to point the diverging behavior more clearly. The dashed and dotted lines locate the pressure CDF when air arrives at the middle of the model and at the filter (breakthrough) for two different system sizes: $w=220$, $l=150$ (dotted red lines) and $w=220$, $l=1330$ (dashed blue lines). As the length of the model increases, the values of $F^*(l/2)$ and $F^*(l)$ tend to converge approaching the critical value $F^*_{crit}$.}
	\label{fig:invasion_sketch}
\end{figure}

In Fig.\,\ref{fig:invasion_sketch} we have plotted in blue (left) the curve $y(F^*)$ obtained by averaging the results from $10$ numerical simulations in a system of width $w=220$ and length $l=546$ to keep the same proportions of the model shown in Fig.\,\ref{fig:evolution}. As anticipated, this curve presents the vertical asymptotic behavior at a given value $F_{crit}^*$, measured to be $F^*_{crit} = 0.52$. In order to verify this divergence more explicitly, we have also plotted the function $y(F^*)$ for a longer system having width $w=220$ and length $l=1330$. The resulting curve is shown as an inset in Fig.\,\ref{fig:invasion_sketch}, where the divergence at $F_{crit}^*=0.52$ can be more clearly seen. We notice that this value is very close to $0.50$, the bond percolation threshold for an infinite square network \citep{stauffer1994}, the difference being possibly due to the fact that the system's width $w=220$ is not too large. We raise the claim that for large enough systems (such that the boundary effects can be neglected), the invasion far from the system's boundaries happens at a statistically constant capillary pressure, whose value can be estimated by

\begin{equation}
p_{crit}=F^{-1}(F^*_{crit}) \:,
\label{eq:pcrit}
\end{equation}
in which $F^*_{crit}$ is the bond percolation threshold \citep{stauffer1994} associated with a network having the same average connectivity as the sample, and $F$ is the cumulative distribution function associated with the values of the capillary pressure thresholds in the sample. Values of the bond percolation threshold are tabulated for several different types of lattices and we present some of them in Table \ref{table}.

\begin{table}
	\centering
		\begin{tabular}{l||l|l}
					Lattice Type & Connectivity & $F^*_{crit}$ \\ \hline
					2D honeycomb & $3$ & $1-2\sin\left(\pi/18\right) \approx 0.6527$ \\
					2D square & $4$ & $1/2 = 0.50$ \\
					2D triangular & $6$ & $2\sin\left(\pi/18\right) \approx 0.3473$ \\
					3D diamond & $4$ & $0.3880$ \\
					3D cubic (simple) & $6$ & $0.2488$ \\
					3D cubic (BCC) & $8$ & $0.1803$ \\
					3D cubic (FCC) & $12$ & $0.1190$ 
		\end{tabular}
		\caption{Bond percolation thresholds for different lattice types in $2$ and $3$ dimensions. The connectivity is understood as the number of nearest neighbors to each pore in the lattice. (adapted from \citep{stauffer1994}).}
		\label{table}
\end{table}

Eq.\,\ref{eq:pcrit} condenses, in a single simplified form, the influence to the critical capillary pressure from several material properties of the porous medium and fluids involved. The pore-size distribution of the medium, wetting properties of the fluids and surface tension between the phases are encoded in the function $F$ and the average pore connectivity is used in the computation of $F^*_{crit}$. This means that knowledge of the fluids' wetting properties and the medium's geometry are sufficient to produce an estimation of the critical pressure $p_{crit}$ prior to any empirical test. Such information could be of much value, for example, for professionals interested in the modeling of two-phase flows in the bulk of large reservoirs (where the influence of the boundaries may be neglected). Additionally, a priori knowledge of the numerical value of $p_{crit}$ could also be of interest for oil recovery, specifically in the so-called secondary recovery stage in which a fluid is injected into the reservoir in order to increase its pressure and drive the oil flow. Overpressurizing a porous medium during drainage may lead to faster flows, with the appearance of viscous effects which can considerably increase the residual saturation of the wetting phase \citep{grunde2011} and reduce the oil recovery. Beforehand estimation of the value of $p_{crit}$ may, therefore, be useful in avoiding such overpressurizing.

To characterize the importance of the system's dimensions in the current analysis, we have used the inverse of Eq.\,\ref{eq:y} to plot on the inset of Fig.\,\ref{fig:invasion_sketch} lines corresponding to the values of $F^*(l)$ and $F^*(l/2)$ in which the invasion front maximum position is $y=l$ (breakthrough) and $y=l/2$ (halfway through the model's length). Two model dimensions were considered: $w=220$, $l=150$ (dotted red lines) and $w=220$, $l=1330$ (dashed blue lines). One can see that, as the length of the model increases, the values of $F^*(l/2)$ and $F^*(l)$ tend to converge approaching the critical value $F^*_{crit}$. It is in this limit that the capillary pressure is essentially constant during the invasion and well estimated by Eq.\,\ref{eq:pcrit}.

In order to verify experimentally our results, we have decided to produce the experimental curve $y(F^*)$ for the invasion shown in Fig.\,\ref{fig:evolution}. To obtain this curve, we have measured the distance $y$ from the images of the flow and used the inverse of Eq.\,\ref{eq:Fstar} to relate the imposed capillary pressure $p_c$ to the experimental CDF value $F^*$. Notice that this requires the experimental determination of the actual capillary pressure thresholds distribution (from which the cumulative distribution $F$ used in Eq.\,\ref{eq:Fstar} is then calculated). Since our image-based method from Sec.\,\ref{sec:voronoi} allows us to find only the distribution of $1/d$, the inverse of pore-throat sizes (see Fig.\,\ref{fig:poresize}), we needed to produce a calibration curve to relate $p_c$ and $1/d$. In an ideal scenario, one would expect from the Young-Laplace law \citep{bear1972} a linear relationship between $p_c$ and $1/d$. In a more realistic case, dynamical effects such as contact angle hysteresis can make such a relationship more complicated. We have estimated this relationship using a separate experiment in which we have employed the same porous medium shown in Fig.\,\ref{fig:evolution}, but this time driven under a constant (slow) withdrawal rate. The choice of this boundary condition was motivated by the fact that, in this case, the oscillations in the measured pressure signal are not as pronounced as in the system driven with the constant imposed pressure, thus yielding a more accurate measurement of the pressure. A linear fit to our data in the region close to critical pressure gives $p_c=4.0\: 10^1\mathrm{Pa.mm}\:(1/d)+3.1\: 10^2\mathrm{Pa}$. Since the majority of the invasion happens at pressures close to this value (in the propagation regime defined previously), this linearization procedure is justifiable in this context.

By using this relationship, we have been able to produce the experimental curve $y(F^*)$, shown in green (right) in Fig.\,\ref{fig:invasion_sketch}. We notice that the divergence of $y$ at a given value of $F^*_{crit}$ is also observed in the experimental case, the critical value measured being $F^*_{crit} = 0.605$. The diverging nature of $y(F^*)$ was expected from the plateau in the experimental \mbox{P-S} curve in Fig.\,\ref{fig:pscpexp3}. The value of $F^*_{crit}$ presented here can only be given up to a precision of $\pm0.060$, due to limitations in both our method of estimating the distribution of $1/d$ (described in Sec.\,\ref{sec:voronoi}) and in the determination of the calibration curve linking $1/d$ and $p_c$. Nevertheless, the difference between the experimental and numerical values of $F^*_{crit}$ is expected because the connectivity of the grid in the simulations is $4$ (square lattice) whereas in the experiments, it assumes an average value between $3$ and $4$, as can be seen from Fig.\,\ref{fig:voronoi}. From Table \ref{table}, we therefore expect a critical value in the interval $0.5<F^*_{crit}<0.6527$, i.e., bounded by the values of $F^*_{crit}$ corresponding to the square lattice (connectivity $4$) and the honeycomb lattice (connectivity $3$).

\section{Conclusions}
In the current work we have analyzed, both numerically and experimentally, the first drainage in a two-phase flow inside a porous medium. We have focused our attention on the influence of boundary effects to the measurement of pressure-saturation curves in such systems and our results indicate that some features of the \mbox{P-S} curves can indeed be highly influenced by these effects. The invasion can be divided into three regimes that, for convenience, we call here building, propagation and clogging regimes.

In the building regime, the air-liquid front evolves from the flat profile (in which the invasion within the porous network starts) to the ramified fractal structure characteristic of flows in the capillary regime. This evolution in the front morphology is responsible for the initial pressure build up in the \mbox{P-S} curve and it happens through a certain distance near the inlet of the system. This distance increases with the size of the inlet boundary itself (the width $w$ in our analysis).

Once the invasion front has attained its ramified fractal structure (typical of slow invasion governed by capillary forces), the front propagates at a capillary pressure fluctuating around a statistically constant value $p_{crit}$. This characterizes the propagation regime, which dominates the dynamics in the limit of large systems, in which the boundary effects are negligible. The particular value of the critical capillary pressure $p_{crit}$ is dependent on material properties of the fluids (such as wettability and surface tension), and the geometry and topology of the medium (via its pore-size distribution and average pore connectivity). Our work establishes a method for directly estimating the value of $p_{crit}$. This is done via Eq.\,\ref{eq:pcrit} and such estimation may be useful for a wide range of practical applications. We particularly stress its possible use in the simulations of two-phase flows in the bulk of large porous networks. As described earlier, \mbox{P-S} curves are used to give closure to the extension of Darcy equations to multiphase flows employed in such simulations.

The clogging regime takes place once the air-liquid interface percolates through the model reaching the filter at the outlet (breakthrough). In the clogging regime, narrower pores in the vicinity of the filter are invaded, leading to the rise of the capillary pressure from the critical level $p_{crit}$ of the propagation regime. This process happens until the clogging of the filter, the moment in which air completely fills the outlet boundary and the liquid in the filter gets disconnected from the porous medium. The invasion of narrower pores in this regime is induced by the presence of the filter itself: since its typical pore-sizes are much smaller than the pore sizes in the model, the air front is constrained to remain inside the porous medium. The final pressure build up is, therefore, a direct result of the artificial placement of a filter at the outlet boundary.

\begin{figure}[t]
	\centering
	\includegraphics[width=\linewidth]{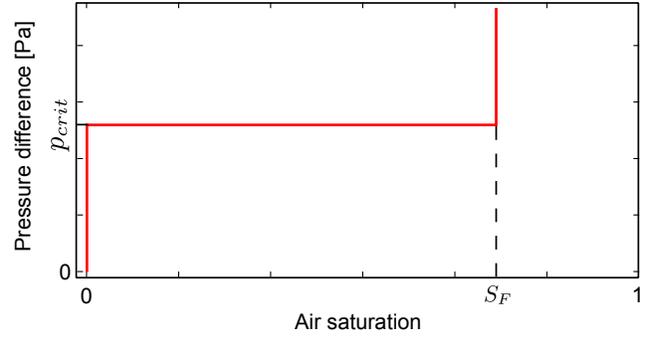}
	\caption{The pressure-saturation relationship for a system unaffected by boundary effects is specified simply by two parameters: a critical capillary pressure $p_{crit}$ and a final air saturation $S_F$.}
	\label{fig:psreduced}
\end{figure}

We emphasize that the boundary effects observed here can also happen in actual measurements of \mbox{P-S} curves (or, alternatively, water-retention curves) in real soil and rock samples. Since the measurements are typically performed inside closed setups, commonly with a filter at the outlet (such as in the porous diaphragm method), boundary effects are reflected in the resulting curves and an analyst using the experimental data must take into consideration the extent to which these effects might influence the resulting analysis. In a realistic invasion pattern inside a sample large enough, none of these boundary effects would be present. In 2D, we believe our much simplified description in terms of two values only, a critical pressure $p_{crit}$ and a final saturation $S_F$, holds a more reliable picture for such scenarios, see Fig.\,\ref{fig:psreduced}. The critical pressure is $p_{crit}=F^{-1}(F^*_{crit})$, Eq.\,\ref{eq:pcrit}, and the final saturation $S_F$ is a random variable with an average value $S_{ave} =\alpha (w/d)^{(D_c-D)}$, Eq.\,\ref{eq:Save}, and a maximum value $S_{max} = \alpha(L/d)^{(D_c-D)}$, Eq.\,\ref{eq:Smax}, for a system of REV of linear size $L$, and (smallest) macroscopic size $w$, where $d$ is the typical distance between adjacent pores.

In 3D systems both phases can percolate simultaneously, contrarily to the two dimensional case \citep{stauffer1994,sahimi2011} where either one or the other percolates. This will result in emptying progressively portions of the defending fluid connected to the outlet after the first one percolates, and even in a large system, make the invasion happen not only at first percolation pressure, but also above -- something closer to the upwards ramps seen in the Brooks-Corey and Van Genuchten models. Nonetheless, this can only happen as long as the defending spanning cluster is present. Once it gets disconnected, only finite size clusters are present, close to the outlet region in a semipermeable system. The invasion of these clusters would lead to a final increase in the capillary pressure -- a finite size effect absent in an open system in 3D. This precise point is interesting to study per se in future research. In a closed 3D system with a filter (still neglecting gravitational effects), we expect that the difference in connectivity properties in percolation (as compared to the 2D case) will yield for the \mbox{P-S} curve a plateau at $p_{crit1}=F^{-1}(F^*_{crit1})$, corresponding to the growth of the percolating invasion cluster, followed by a rising function of saturation, going from $p_{crit1}=F^{-1}(F^*_{crit1})$ to $p_{crit2}=F^{-1}(F^*_{crit2})$, where $F^*_{crit1}$ is the percolation threshold for the invading phase coming in, and $F^*_{crit2}$ (that should be $1-F^*_{crit1}$) is the percolation threshold of the defending phase at residual saturation. Once $p_{crit2}$ is overcome, the large size defending cluster disconnects, and for large systems the saturation does not change significantly anymore if the pressure difference is increased beyond $p_{crit2}$. Hence, the modification from the 2D behavior, for large open system, is the addition of a rising ramp for the \mbox{P-S} curve from $p_{crit1}$ to $p_{crit2}$ between the horizontal and final vertical asymptotes of Fig.\,\ref{fig:psreduced}.

Additional effects that were not studied in the present work can also play an important role in the determination of \mbox{P-S} curves. For example, flow through corners in the porous medium \citep{lenormand1984,tuller2001} may increase the hydraulic continuity of the wetting phase, providing a pathway for the long-term drainage of clusters that would be otherwise completely trapped in the middle of the non-wetting phase. The extent of the region connected by corner flow depends on specific properties of the porous medium and fluids involved and this effect is particularly important in the case of angular pore spaces \citep{tuller1999}. For the system studied in the present work, corner flow is not so relevant due to the fact that the ``corners'' in which the wetting phase can accumulate correspond to the isolated points where the glass beads touch the upper and lower contact paper surfaces. Since these points are spatially localized, the formation of long pathways of regions hydraulically connected via corner flow is unlikely. In the case of tests to determine the \mbox{P-S} curves using real porous media, corner flow may be important depending on the particular porous medium and fluids employed. Nevertheless, if the sample is long enough, these corners or films could rupture/disconnect over large length and time (depending on the specific porous material and fluid pairs, for the many cases where none of the fluids is completely wetting the solid components). Therefore, the extent of the region of pores hydraulically connected to the outlet via corner flow could become negligible in comparison to the sample size and such effects can then also be seen as boundary effects associated to the flow close to the outlet. In this limit of a large enough sample, the results derived in this paper should remain applicable.

\appendix
\section{Description of the overflowing mechanism}\label{apxA}
\begin{figure}[t]
	\centering
	\includegraphics[width=\linewidth]{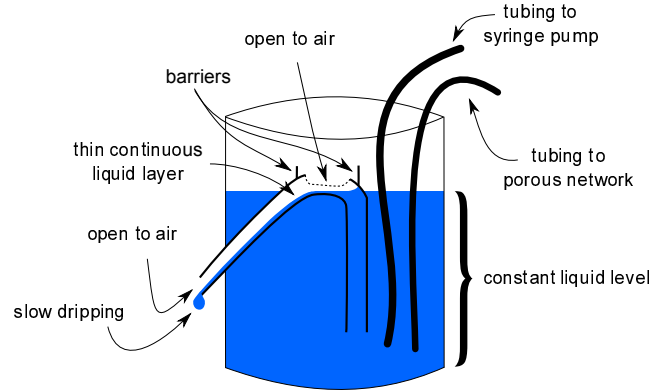}
	\caption{Diagram of the overflowing mechanism with the siphon-like tube constructed in the liquid reservoir. The width of the continuous liquid layer inside the inclined tube and the size of the droplet dripping from the left are both exaggerated for better visualization.}
	\label{fig:diagram_bottle}
\end{figure}

Fig.\,\ref{fig:diagram_bottle} shows a detailed diagram of the overflowing mechanism built into the fluid reservoir used to control the imposed pressure in the experiments. A siphon-like tube is built with one inclined end going out from the side of the reservoir and a vertical end hanging inside of it. As the reservoir is filled with liquid, the vertical part of the tube is also filled while the inclined one remains with air. When the liquid level reaches a certain height, a thin liquid layer is formed on the top of the curved part of the tube and is guided through the inclined side to finally drip slowly outside of the reservoir. A hole is cut from the extreme top of the tube (dashed line in the diagram) in order to avoid the meniscus inside of it (on the right side of the hole in the diagram) to reach the other side, which could eventually lead to completely flushing out the liquid from the bottle (like an actual siphon). A small vertical barrier is built around the hole to avoid liquid from the reservoir to go into it. One tube connects the reservoir to the porous medium while another one connects it to an external syringe pump. Apart from the tubing, the whole structure is made of glass. Liquid is constantly pumped into this reservoir from the external syringe pump at a very low rate. The purpose of this inflow is two-folded: on one hand it overcomes the losses due to evaporation, on the other it ensures the stability and continuity of the thin liquid layer inside the siphon-like tube, avoiding any sort of intermittency effect in the meniscus inside of it. The intermittency mentioned is of the same kind as the one observed when one tries to slowly fill a glass with water up to the top. Before spilling, a meniscus is formed on the top of the glass, which grows up to a maximum size and then bursts, spilling the water out of the glass. This sudden behavior could cause abrupt changes in the height level of the order of $1mm$ introducing an error larger than $10Pa$ in the imposed pressure. By guaranteeing the existence of the continuous liquid layer inside the curved tube, this problem is also avoided. The design of this reservoir is based on a modification of a very curious object called a Tantalus cup (or Pythagorean cup), which we highly recommend the interested reader to search for.

\section{Further analysis of the air saturation}\label{apxB}
\subsection{Variation of the final air saturation with the system's width}\label{apx1}

\begin{figure}
\noindent\includegraphics[width=20pc]{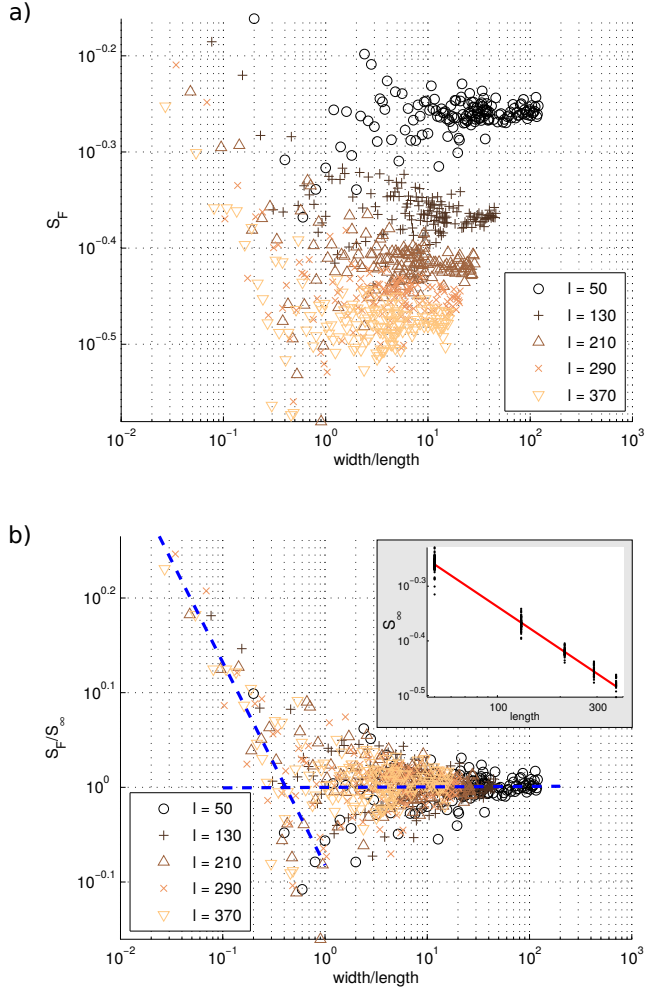}
\caption{Simulations. a) Final air saturations as a function of $w/l$ for several lengths with $w$ varying from 10 to 6000. b) Same curves divided by $S_{\infty}$, the average value of of $S_F$ at large $\frac{w}{l}$. The dashed blue line corresponds to the linear regression giving the exponent $D_c\simeq1.79\pm0.10$. The inset corresponds to $S_{\infty} \propto l^{(D_c^\infty-D)}$ with $D_c^\infty\simeq1.74\pm0.10$ and $D=2$.}
\label{fig:width_satFin}
\end{figure}

In this appendix section we extend the analysis initiated in Sec.\,\ref{sec:sat} by considering the variation of the final air saturation as the width of the system is gradually changed. Fig.\,\ref{fig:width_satFin}a) shows the evolution of $S_F$ with $r'=w/l$ (inverse of $r=l/w$ used in Sec.\,\ref{sec:sat}) when increasing $w$ from $w=10$ to $6000$ with a step of 10 from 10 to 200, 20 from 200 to 2000 and 100 from 2000 to 6000. The procedure is repeated for several different lengths $l$ (shown in the legend in the figure). After rescaling the data by dividing it by $S_{\infty}(l)$, i.e., the value of the final saturation for a system with given length $l$ in the limit of very large width, we obtain the data shown in Fig.\,\ref{fig:width_satFin}b). Once again we postulate the scaling

\begin{equation}
S_F(r',l)=r'^{(D_c-D)}h(r')S_{\infty}(l) \:,
\label{eq:Sf2}
\end{equation}
where $h(r')$ is a crossover function, being defined similarly to Eq.\,\ref{eq:f}:
\begin{equation}
h(r') =
\left\{
	\begin{array}{ll}
		C'  & \mbox{for small } r' \\
		r'^{-(D_c-D)} & \mbox{for large } r',
	\end{array}
\right.
\label{eq:h}
\end{equation}
where $C'$ is again a dimensionless constant, numerically determined as $C'\simeq 0.83$ by a linear regression in the bilogarithmic plot shown on Fig.\,\ref{fig:width_satFin}b), over the ensemble of simulations at $r'<0.4$. From the slope of the curve fitting the small $r'$ data in Fig.\,\ref{fig:width_satFin}b) we estimate the fractal dimension to be again $D_c\simeq1.79\pm0.10$. Additionally, $S_{\infty}(l)$ is found to scale as $S_{\infty} \propto l^{(D_c^\infty-D)}$ with $D_c^\infty\simeq1.74\pm0.10$ and $D=2$ as before, see the inset in Fig.\,\ref{fig:width_satFin}b). The turning point for the crossover function happens approximately at $r'=0.4$, i.e., when the length is about $2.5$ times larger than the width. We would like to remark that the convergence to a flat plateau is expected, once again, due to the same symmetry arguments presented before. The only difference in the present case, being the fact that the statistical symmetry for the final saturation arises in the direction perpendicular to the inlet-outlet direction. If we consider two subwindows one next to the other in this direction, for large enough systems (large $w$), there is no reason to believe that the final saturation measured in one of these subwindows would be different than the same quantity measured in the other (which is also the same as the saturation measured in a subwindow formed by the union of the two). This kind of statistical symmetry holds for the central region of the system, in which the effects from the side boundaries are negligible. As the width of the system is increased, this region grows and becomes dominant for large enough systems.

\subsection{Variation of the breakthrough saturation with the system's length}\label{apx2}
\begin{figure}
\noindent\includegraphics[width=20pc]{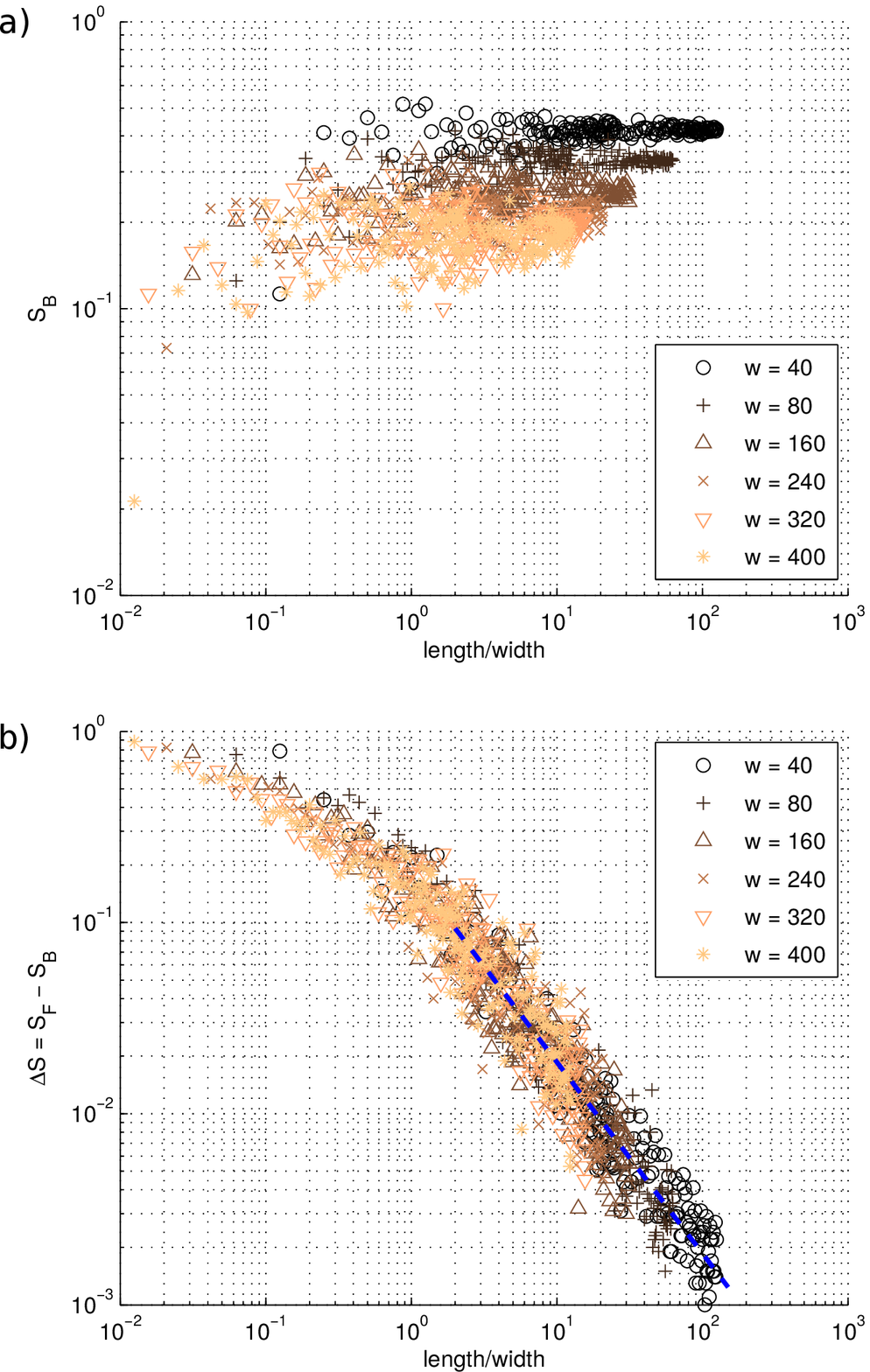}
\caption{Simulations. a) Breakthrough air saturations. b) Difference between the final and the breakthrough air saturations. The dashed blue line shows the scaling $\Delta S \propto\left(l/w\right)^{-1}$.}
\label{fig:length_satTh}
\end{figure}

We recall that the breakthrough saturation $S_B$ is defined as the saturation at the moment when air percolates for the first time, reaching the exit side and forming a connected sample-spanning cluster of invaded pores ($S_B<S_F$). We have plotted the breakthrough saturation $S_B$ (Fig.\,\ref{fig:length_satTh}a)) and the difference $\Delta S = S_F - S_B$ (Fig.\,\ref{fig:length_satTh}b)) as functions of the ratio $r=l/w$, where again different colors and symbols are used for different values of $w$. The values of $S_B$ at fixed $w$ in Fig.\,\ref{fig:length_satTh}a) are essentially constant, with some dispersion for systems with very small lengths. Yet again, the existence of the plateau is due to the translational invariance along the inlet-outlet direction described in Sec.\,\ref{sec:sat_l}. In Fig.\,\ref{fig:length_satTh}b), the decrease of $\Delta S$  indicates that final and breakthrough saturation are only significantly different for small systems. This is also intuitively expected. The breakthrough saturation is measured at the moment in which air first percolates through the medium, reaching the filter at the outlet. After this instant, the pores in the vicinity of the filter start to be invaded, until the moment in which the filter gets clogged (which in the simulations corresponds to the moment in which the last line of pores is completely filled with air). Therefore, the difference $\Delta S$ between the values of $S_F$ and $S_B$ is due to the invasion of those last pores in the vicinity of the filter. The number of pores in this region scales as $n_{filt} \propto w^2$ while the total number of pores in the system is proportional to the area of the porous matrix, i.e., $n_{tot} \propto wl$. In the limit of long systems, $\Delta S$ scales as the ratio $n_{filt}/n_{tot}$, therefore, $\Delta S \propto\left(l/w\right)^{-1}$. This scaling is shown in Fig.\,\ref{fig:length_satTh}b) by the dashed blue line. As the system's length gets longer, the fraction of pores in the region close to the filter becomes smaller as compared to the total number of pores, thus giving a minor contribution to the saturation. For such large systems, $S_F$ is still larger than $S_B$, but not much. Conversely, for small $l$, the number of pores in the vicinity of the filter becomes relatively large in comparison to the total amount of pores, thus the larger values for $\Delta S$.

\subsection{Variation of the breakthrough saturation with the system's width}\label{apx3}
\begin{figure}
\noindent\includegraphics[width=20pc]{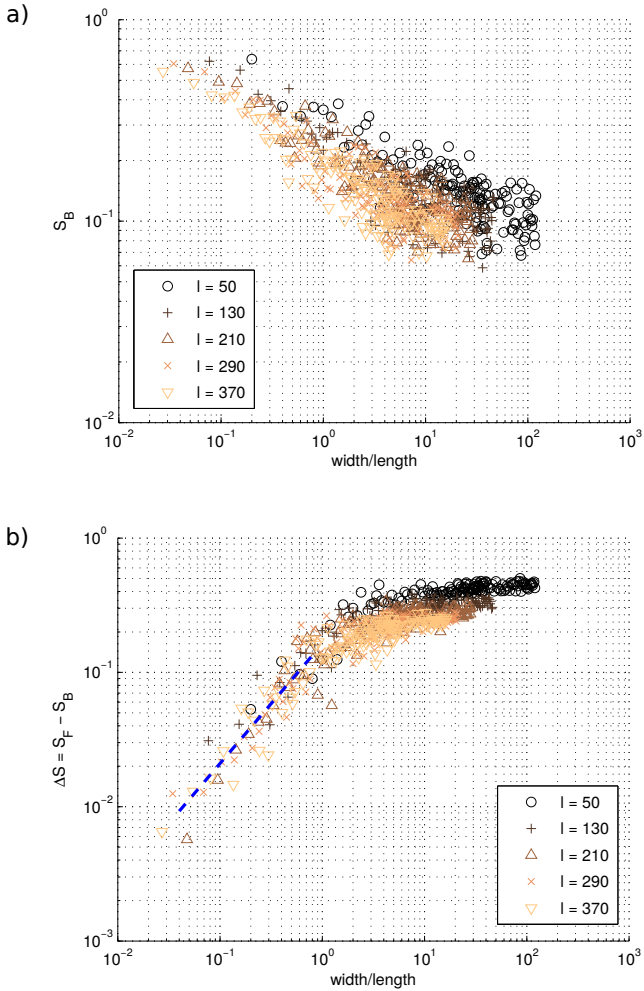}
\caption{Simulations. a) Breakthrough air saturations. b) Difference between the final and the breakthrough air saturations. The dashed blue line shows the scaling $\Delta S \propto\left(w/l\right)^{1}$.}
\label{fig:width_satTh}
\end{figure}

Fig.\,\ref{fig:width_satTh}a) shows that the value of $S_B$ is reduced as the system gets wider. There is an intuitive reasoning behind this behavior: the larger the system gets, the more probable it is for an easy path to exist (consisting of pores with low capillary threshold values) connecting the inlet to the outlet. The breakthrough can thus be reached at an earlier stage of invasion, leading to a lower value of $S_B$. Fig.\,\ref{fig:width_satTh}b) shows the behavior of the difference $\Delta S = S_F - S_B$ with the ratio $r'=w/l$. The dashed blue line corresponds to the same scaling discussed in the previous section. Since the abscissa here is $w/l$, the exponent changes sign and $\Delta S \propto\left(w/l\right)^{1}$. The increase in this quantity as the system gets wider is due to the decrease in $S_B$ with the width, as discussed.

\section{Comparison between the image-based estimation of the capillary pressure thresholds distribution for different models}\label{apxC}
In Fig.\,\ref{fig:pscomp3} we have noticed that the capillary pressure driving the invasion for the three models studied differed a bit from one model to another. We have argued that this difference came from the fact that the models were rebuilt for each experiment and therefore the pore-size distribution varied from one to another, which could then result in the observed differences. In order to clarify this point, we have performed the image-based estimation of the capillary pressure thresholds distribution for the other models, in a similar manner to what had been done previously for the longest model (see Sec.\,\ref{sec:voronoi}).

\begin{figure}
\noindent\includegraphics[width=20pc]{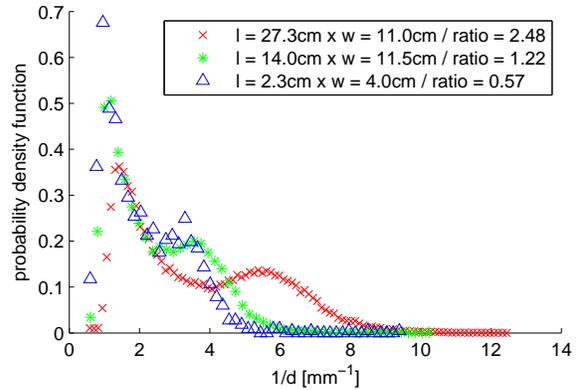}
\caption{Image-based estimation of the capillary pressure thresholds distribution for the three models studied experimentally. The model's geometries are given in the legend. We observe that the longest model has narrower pores (higher values of $1/d$), which is reflected in the higher value of the entrance pressure seen in Fig.\,\ref{fig:pscomp3}.}
\label{fig:voronoicomparison}
\end{figure}

Fig.\,\ref{fig:voronoicomparison} shows the resulting distributions of $1/d$, the inverse of the pore-throat size, which gives an estimation of the capillary pressure according to Young-Laplace law as described in Sec.\,\ref{sec:voronoi}. The colors in Fig.\,\ref{fig:voronoicomparison} were chosen as to match the ones in Fig.\,\ref{fig:pscomp3}, different colors for different models. The models' geometries are given in the legend of the figure. The distribution shown for the longest model (red crosses data) is the same as the one in Fig.\,\ref{fig:poresize}. We see that the smallest model (blue triangles data) has lower values of $1/d$ (wider pore-throats), which in its turn is reflected in the lower capillary pressure needed to drive the invasion, as observed in Fig.\,\ref{fig:pscomp3}. The model having intermediate size (green stars data) has a distribution lying in between the smallest and longest ones as shown in Fig.\,\ref{fig:voronoicomparison}. Notice that, the fact that the critical pressure $p_{crit}$ seems to be increasing with model size in Fig.\,\ref{fig:voronoicomparison} is a coincidence: the difference in the values of capillary pressure thresholds (and consequantly $p_{crit}$) depend on the pore-throat size distribution but not necessarily on the sample dimensions.

\begin{acknowledgments}
The first and second authors contributed equally to this work. We would like to thank the support from the University of Oslo, University of Strasbourg, the Norwegian Research Council through the FRINAT Grant No. 205486, the REALISE program and the EU Marie Curie ITN FLOWTRANS network. This project has also received funding from the European Union's seventh framework programme for research, technological development and demonstration, under grant agreement No. 316889, ITN FlowTrans. We acknowledge the support of the IDEX program through the award ``Hope of the University of Strasbourg'', and the University of Oslo through a guest researcher programme. Additionally, we thank Alex Hansen, Eirik Flekk\o{}y, Rubens Juanes, Benjy Marks, Ren\'e Castberg and Mihailo Jankov for valuable discussions and technical support. The data used in this paper can be obtained upon request.
\end{acknowledgments}

\end{article}

\end{document}